\documentclass[preprint,12pt]{elsarticle}

\usepackage{graphicx}
\usepackage{amssymb}
\usepackage{amsmath}
\usepackage{hyperref}
\usepackage{xcolor}

\begin{document}

\begin{frontmatter}

\title{Efficient Self-Supervised Barlow Twins from Limited Tissue Slide Cohorts for Colonic Pathology Diagnostics} 
\author[1]{Cassandre Notton}
\author[1]{Vasudev Sharma}
\author[3]{Vincent Quoc-Huy  Trinh}
\author[4]{Lina Chen}
\author[5]{Minqi Xu}
\author[6]{Sonal Varma}
\author[1,2]{Mahdi S. Hosseini}

\affiliation[1]{organization={University of Concordia},
            addressline={1455 De Maisonneuve Blvd. W.}, 
            city={Montreal},
            postcode={H3G 1M8}, 
            state={Quebec},
            country={Canada}}
						
\affiliation[2]{organization={Mila--Quebec Artificial Intelligence Institute},
            addressline={6666, St-Urbain, \#200}, 
            city={Montreal},
            postcode={H2S 3H1}, 
            state={Quebec},
            country={Canada}}
						
\affiliation[3]{organization={University of Montreal},
            addressline={2900, boul. Edouard-Montpetit}, 
            city={Montreal},
            postcode={K3T 1J4}, 
            state={Quebec},
            country={Canada}}

\affiliation[4]{organization={Sunnybrook Health Science Centre},
            addressline={2075 Bayview Avenue}, 
            city={Toronto},
            postcode={M4N 3M5}, 
            state={Ontario},
            country={Canada}}
            
\affiliation[5]{organization={Kingston General Hospital},
            addressline={76 Stuart Street}, 
            city={Kingston},
            postcode={K7L 2V7}, 
            state={Ontario},
            country={Canada}}

\begin{abstract}
Colorectal cancer (CRC) is one of the few cancers that have an established dysplasia-carcinoma sequence that benefits from screening. Everyone over 50 years of age in Canada is eligible for CRC screening. About 20\% of those people will undergo a biopsy for a pre-neoplastic polyp and, in many cases, multiple polyps. As such, these polyp biopsies make up the bulk of a pathologist's workload. Developing an efficient computational model to help screen these polyp biopsies can improve the pathologist's workflow and help guide their attention to critical areas on the slide.  DL models face significant challenges in computational pathology (CPath) because of the gigapixel image size of whole-slide images and the scarcity of detailed annotated datasets. It is, therefore, crucial to leverage self-supervised learning (SSL) methods to alleviate the burden and cost of data annotation. However, current research lacks methods to apply SSL frameworks to analyze pathology data effectively. This paper aims to propose an optimized Barlow Twins framework for colorectal polyps screening. We adapt its hyperparameters, augmentation strategy and encoder to the specificity of the pathology data to enhance performance. Additionally, we investigate the best Field of View (FoV) for colorectal polyps screening and propose a new benchmark dataset for CRC screening, made of four types of colorectal polyps and normal tissue, by performing downstream tasking on MHIST and NCT-CRC-7K datasets. Furthermore, we show that the SSL representations are more meaningful and qualitative than the supervised ones and that Barlow Twins benefits from the Swin Transformer when applied to pathology data. Codes are avaialble from \url{https://github.com/AtlasAnalyticsLab/PathBT}.
\end{abstract}

    
\end{frontmatter}

\listoffigures
\listoftables
\newpage

\section{Introduction}
Computational Pathology (CPath) is an innovative field at the intersection of pathology and computer science. It aims to successfully create a framework of digital diagnostics that helps extract meaningful representations from the raw data in the oncology domain.  Leveraging tools such as Deep Learning (DL), CPath enhances histopathological Whole Slide Images (WSIs) analysis and provides valuable diagnosis insights \cite{hosseini2024computational}. CPath is expected to integrate easily with existing diagnostic workflows, improving diagnostic accuracy, reproducibility, and disease detection and grading efficiency \cite{kather2020development}. The recent advancements in DL methods in CPath have led to enhanced performance, but new challenges have emerged.

DL methods in CPath often require large amounts of annotated data. However, annotating data is a time-consuming and expensive task that requires the expertise of pathologists with years of extensive clinical experience. This leads to a scarcity of publicly available, detailed annotated datasets and the need to leverage these limited annotations to train accurate models. As digital pathology expands worldwide, identifying methods less dependent on costly annotations becomes critical. To exploit large unlabeled or weakly annotated datasets, it is crucial to capitalize on Self-Supervised Learning (SSL), which trains on large amounts of unlabeled data and can outperform supervised pre-training \cite{DINO}. Additionally, recent works have shown that SSL pre-training on pathology data can improve performance on downstream pathology tasks \cite{Benchmark}. This paper focuses on addressing DL challenges in CPath to classify colorectal polyps and prevent colorectal cancer (CRC). CRC stands as one of the most prevalent causes of cancer-related mortality, but it is also one of the most preventable cancers. However, bottlenecks in patient screening schedules due to a shortage of pathologists induce delayed diagnoses \cite{patho-shortage}. Therefore, integrating CPath into clinical workflows could enhance diagnosis and lead to rapid care, increasing survival as deadly diseases, notably cancer, are detected precisely and efficiently. Hence, developing efficient SSL screening tools for accurately classifying colorectal polyps is crucial for effective CRC screening \cite{Current-GI-cancer}.

Integrated into Gastrointestinal (GI) cancer screening, CPath has shown excellent diagnosis accuracy, on par with board-certified pathologists. With the need for precise cancer grading for treatment personalization, DL is now a cornerstone of the fight against cancer and has achieved high Area Under the Curve (AUC) scores above 0.98. Additionally, it was shown that the deepest models obtain the highest performance due to their superior generalization capabilities \cite{Current-GI-cancer,hosseini2024computational}. 

However, such models require more extensive training data \cite{Current-GI-cancer}. Pathology images are highly specific as they have neither canonical orientations nor high colour variations. The Field of View (FoV) of the patches used for model training is also significant, as different FoV convey varying contextual information \cite{Benchmark}. Despite advancements in GI and CRC screening, current methods do not adequately study the appropriate FoV for CRC screening. Additionally, they lack a thorough comparison of features learned through supervised and self-supervised learning (SSL) approaches. Furthermore, a more comprehensive analysis is needed on how self-supervised learning methods, originally developed for natural images, can be adapted to the unique characteristics of pathological data, particularly through data augmentation techniques.

The future of CPath relies on SSL and its adaptation to the pathology data's specific characteristics. In this paper, we conduct an in-depth analysis of the Barlow Twins \cite{BT} framework and propose to optimize this SSL framework for feature embedding from colorectal polyps for CRC screening.

Our study yields several contributions: we propose (1) an enhanced Barlow Twins framework for pathology data by adapting the hyperparameters, augmentation strategy and pretraining strategy, (2) an evaluation of Barlow Twins representations on the patch and slide level introducing this SSL method in a MIL framework, (3) an investigation on the best FoV for CRC screening, (4) quantitative and qualitative comparisons of the SSL and supervised features, and (5) a new benchmark dataset for CRC screening after transferring the weights from our private dataset to MHIST and NCT-CRC-7k \cite{Clam,mhist,crc-dataset}. Code and pre-trained model weights (in PyTorch) are available at \href{https://github.com/AtlasAnalyticsLab/PathBT}{https://github.com/AtlasAnalyticsLab/PathBT} for further contributions to the research community.

The paper is organized as follows: Section \ref{sec_related} reviews the related work and discusses the existing approaches to self-supervised learning in CPath. Section \ref{sec_methods} describes the proposed methodology, detailing the patch dataset curation and the different techniques used. Section \ref{sec_results} presents the experimental results and analyzes the qualitative and quantitative performance of the proposed methods, including heatmap and UMAP examination. In Section \ref{sec_ccl}, we discuss the implications of our findings and suggest potential directions for future research.

\section{Related Work}
\label{sec_related}

\subsection{Computational pathology}
The development of rapid and efficient whole-slide imaging technology has laid the groundwork for CPath. This discipline leverages computational methods to analyze and characterize histopathological features within gigapixel digital images, or Whole Slide Images (WSIs), generated from biomedical microscopy. CPath aims at improving cancer diagnosis, prognosis, and treatment planning \cite{hosseini2024computational}. The introduction of Deep Learning techniques for WSI representation learning has significantly enhanced diagnostic performance, achieving results comparable to those of board-certified pathologists \cite{hou2016patch}.
\subsection{Self-Supervised Representation Learning}
Self-Supervised Learning (SSL) methods learn representations of a training dataset through pre-text tasks, or SSL paradigms. Nowadays, three of them are particularly present in the literature. On the one hand, \textbf{Contrastive Learning} (CL) methods, such as SimCLR \cite{chen2020simple} or MoCO \cite{he2020momentum}, have become increasingly successful. These methods consist of learning representations that attract distorted views of the same input and push apart the representations of different inputs. On the other hand, \textbf{non-contrastive methods}, such as Barlow Twins \cite{BT}, are also built on the objective of learning similar representations for distorted views of the same input, but representations from different inputs are not used in the process. Barlow Twins is designed to learn invariant representations of the data under different input distortions. This framework passes two batches of distorted views through an encoder (ResNet-50) and a projector. The embeddings from both distorted batches are used to calculate the cross-correlation matrix. The core objective of this framework is to minimize the redundancy in these representations, and this is achieved by pushing the cross-correlation matrix closer to the identity matrix through the Barlow Twins loss: \begin{equation} \label{eq:l_bt}
    L_{BT}=\sum_i(1-C_{ii})^2+\lambda \sum_i\sum_{j\neq i}C_{ij}^2
\end{equation} 

Minimizing redundancy aligns with the \textit{redundancy-reduction principle} of H. Barlow, suggesting that the brain efficiently processes visual information by discarding redundant features across the visual neural pathway. These two methods rely on a Convolutional Neural Network (CNN) encoder, such as ResNet-50 \cite{ResNET50}. 

Besides, the effectiveness of Vision Transformers (ViT) has been demonstrated on various computer vision tasks \cite{dosovitskiy2021image, vaswani2023attention}. \textbf{DINO framework} has introduced the possibility of training ViT in a Self-Supervised manner \cite{DINO}. 
\subsection{Self-Supervised Learning for pathology data}
As the annotation of large cohorts of data is significantly time-consuming for pathologists, the development of SSL methods for unlabeled pathology data has become increasingly crucial. While various studies in digital pathology have explored the use of contrastive learning for acquiring robust representations for classification tasks \cite{SSL-WSI,ContrastiveSSL}, recent studies suggest limitations to its efficiency for histopathology image analysis \cite{ClusterConstraints}. The limitation arises from the inherent similarity in morphological features between adjacent patches within a WSI, rendering them less effective negative pairs for CL. This is one of the reasons why the DINO framework remains a popular choice for pathology representation learning (HIPT, Clam, Vim4Path \cite{HIPT,Clam,nasiri2024vim4path}). 

A recent benchmark study compared four SSL methods, among which MoCov2, Barlow Twins and DINO, on four pathology datasets \cite{Benchmark}. This study highlighted the need for domain-aligned pretraining with different augmentation techniques such as random vertical flips, weak colour jittering, RandStainNA, and multiple Fields of View. The authors emphasize the potential of domain-specific SSL methods to improve the performance of models fine-tuned for downstream pathology tasks. A study investigating augmentation techniques within the MoCov2 framework for Chest X-rays highlights the need for future research on the impact of various data augmentations as well as their combination for SSL in pathology \cite{pmlr-v143-sowrirajan21a}. A recent work proposes a rotation method on the patch level to apply random rotation to one patch or a crop of that patch and show great results \cite{alfasly2023rotationagnostic}. This emphasis underscores the critical role of SSL frameworks in digital pathology, particularly when complemented by domain-aligned fine-tuning. While prior works, such as \cite{maleki2024self}, which introduces automatic adversarial style augmentation (AdvStyle) to simulate covariate shifts similar to staining variations in pathology data, and \cite{vray2024distill}, which explores diverse augmentation techniques for pathology images within the DINO framework, have made significant contributions, have made significant contributions, no published research to date has investigated fine-tuning Barlow Twins framework for pathology downstream tasks.

Barlow Twins has performed excellently on diverse pathology datasets \cite{Benchmark} and requires less computational resources than DINO \cite{DINO}: it does not require a large batch size and only pushes one encoder on GPUs. Therefore, in this work, we study the hyperparameters and augmentation techniques of this SSL method for colorectal polyps screening tasks.


\subsection{Multiple Instance Learning for WSIs}
Pathologists use different resolutions of the image to dress their final diagnosis. Therefore, Multi Instance Learning (MIL) consists in leveraging the hierarchy of the data to learn a global representation of a WSI or a patient. In such processes, the training is done by aggregating different levels of the data. In digital pathology, Campanella et al. propose an end-to-end weak-supervision framework using a RNN for slide aggregation \cite{Campanella}. Lu et al. show that MIL can leverage low-quality representations for weakly-supervised slide-level tasks \cite{lu2021data}. Following this work, SSL frameworks such as SimCLR, Moco or DINO have been used as instance-level feature extractions. Recent works such as DeepMIL, followed by DeepSMILE, introduce the use of attention pooling methods to weigh the patches by importance for classification \cite{DeepMIL,DeepSMILE}. The authors demonstrate that attention-based MIL outperform regular MIL. More recently, the SOTA method CLAM uses attention to identify subregions of high diagnostic value within a WSI \cite{Clam}. Following recent wors \cite{mammadov2024selfsupervision,vim4path} employing CLAM to compare different encoders, we leverage CLAM as a MIL framework as well to compare the different architectures on the slide level.
\section{Methods}
\label{sec_methods}
\subsection{Benchmark dataset}
\label{sec_data}
\begin{figure*}[ht]
\begin{center}
  \includegraphics[scale=0.3]{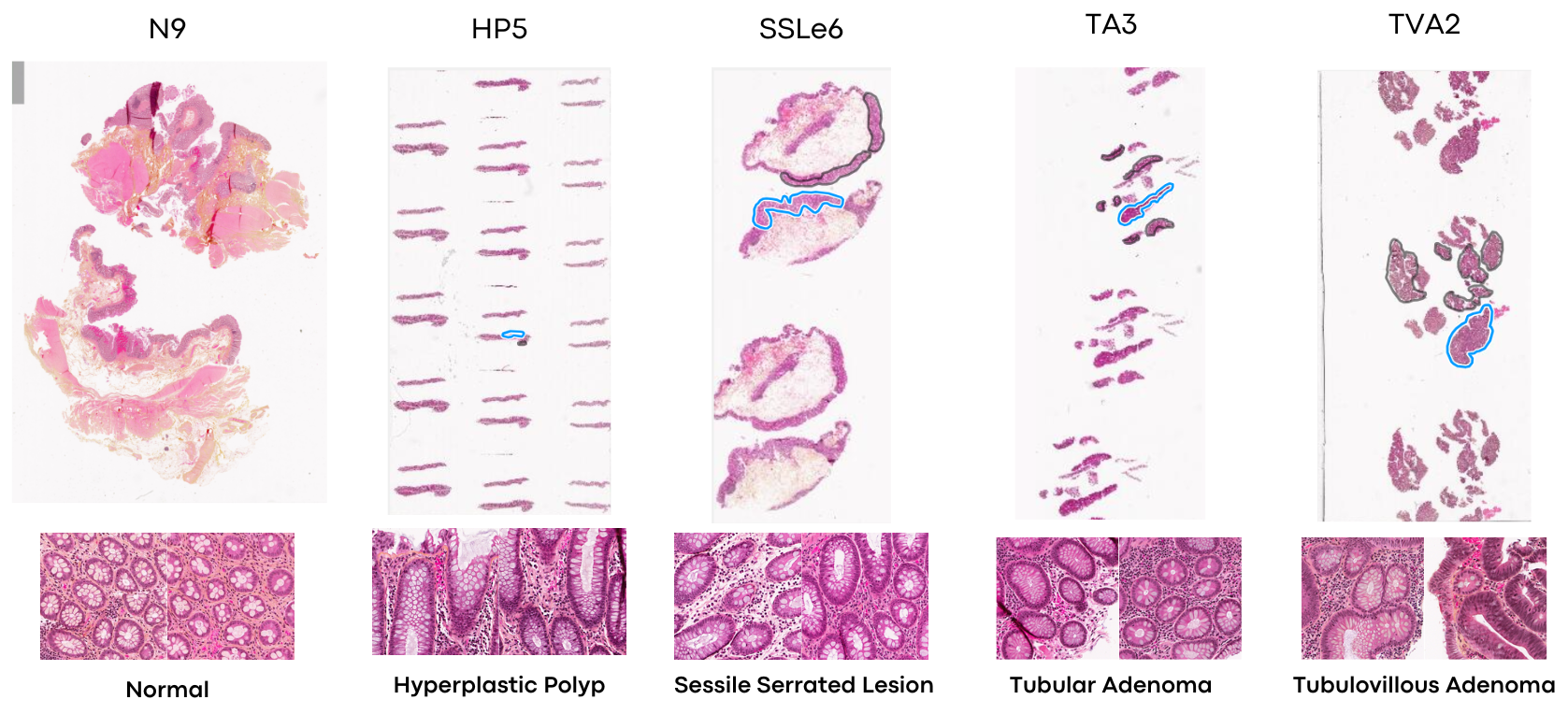}
\end{center}
  \caption{Five slides from the five classes with their annotations (in black and blue) and patches from the ROI for the four polyps and from tissue regions for the Normal WSI. We observe that the annotations are not complete, as only one layer of the sample has been annotated (HP5, SSL6, TA3 and TVA2). The normal slide N9 does not present with any annotations.}
  \label{fig_slides_rois}
\end{figure*}

Our work aims to answer a clinical problem defined by pathologists: the need to develop a screening method for colorectal polyps to recognize and classify different types of polyps. Indeed, this work exploits a Kingston General Hospital (KGH) dataset comprising 1037 WSIs of normal colon tissue and four different types of colorectal polyps: Hyperplastic Polyps (HP), Sessile Serrated Lesions (SSLe), Tubular Adenoma (TA) and Tubulovillous Adenoma (TVA). An expert pathologist partially annotated this dataset on the Region of Interest (ROI) level. A description of these precancerous lesions is provided in Section \ref{sec_ab_data}. The WSIs have been acquired under 4 different magnification levels: 20X, 5X, 1.25X and 0.3125X, with a pixel resolution at 20X of 0.4 microns per pixel (mpp). Table \ref{tab_KGH} presents the slides and annotations throughout the dataset, and Figure \ref{fig_slides_rois} presents five slides from the five classes with the annotations and patches from the ROI when applicable. 
\begin{table}[htp]
    \centering
    \small
    \begin{tabular}{|c|c|c|}
    \hline
        \textbf{Class} & \textbf{\# of slides} & \textbf{\# of ROIs} \\ \hline \hline
        Normal & 200 & 0 \\ \hline 
        TA & 207 & 465 \\ \hline
        SSLes & 201 & 548 \\ \hline
        HP & 212 & 284 \\ \hline
        TVA & 217 & 842 \\ \hline
    \end{tabular}
    \caption{Dataset statistics: slides and annotations in KGH dataset}
    \label{tab_KGH}
\end{table}

For a better understanding of the annotations, we denote $Ann$ the set of annotated tissue in the ROIs, $nonAnn$ the set of tissue non-annotated in the pathological slides, $N$ the tissue from the normal WSIs and $P$ the pathological tissues. We highlight that:
\begin{itemize}
    \item $Ann \subset P$ and $Ann \cap N = \emptyset$ as all ROI regions identify pathological tissues either from HP, SSLes, TA or TVA.
    \item $nonAnn \subset (N \cup P)$ as not all pathological regions have been annotated and regions outside ROIs could be either normal tissue, pathological tissue, or distorted tissue.
\end{itemize}

Distorted tissue is an abnormal alteration in the histological structure, organization, or architecture of tissue caused by tumours, inflammation, medical procedures, or other diseases.

Therefore, we define our dataset as a weakly annotated dataset.
We reserve 100 slides for testing purposes. We extract patches throughout all WSIs using TIAToolBox \citep{Pocock2022} and a custom MLP to remove artifacts and background under four different fields of view: 1400 microns, 800 microns, 600 microns and 410 microns. These datasets will be referred as pkgh, pkgh-800, pkgh-600 and pkgh-410. Figure \ref{fig_patches} shows the different pathologies at different resolutions. Figure \ref{fig_nb_patches} presents the number of patches, within and outside the ROI annotations, for train and test WSIs. It is important to note that using a smaller field of view results in more image patches; however, the same quantity of tissue is represented throughout each dataset.

This work leverages the KGH dataset to optimize Barlow Twins architecture. This approach aims to achieve accurate classification at both the patch and slide levels. Moreover, this approach will be evaluated on PCam dataset \cite{pcam}, a well-known challenging benchmark for classifying metastatic cancer in breast cancer patients' lymph nodes.

\subsection{Enhanced Barlow Twins methods}
This work aims to fine-tune Barlow Twins \citep{BT} for pathology downstream tasks. 

We initially conducted ablation studies on a representative subset of the dataset pkgh-410, made of 5,000 patches per class from ROI to gain insights into the framework's behaviour and identify potential improvements. 
\subsubsection{Barlow Twins hyperparameters}
We first led an \textbf{ablation study on the main hyperparameters} of this framework: the batch size, the projector dimension and the trade-off parameter $\lambda$. The details are shown in Section \ref{sec_ab_hp_transfo}. Our initial ablation study revealed a positive correlation between batch size and performance. However, this improvement comes at the cost of increased computational demands. Therefore, a pragmatic choice of a batch size of 512 was made. Additionally, the default value of the hyperparameter $\lambda$ was adapted to this specific dataset. The ablation study investigating projector dimension yielded a surprising result. While the original paper suggested a direct relationship between accuracy and projector dimension, our findings indicated that a dimension of 2048 was optimal for this dataset, suggesting that the representations from the polyps overfit in a larger dimensional space. 
\subsubsection{Barlow Twins augmentation strategy}
\label{sec_aug}
\begin{figure}[htp]
\begin{center}
  \includegraphics[scale=0.35]{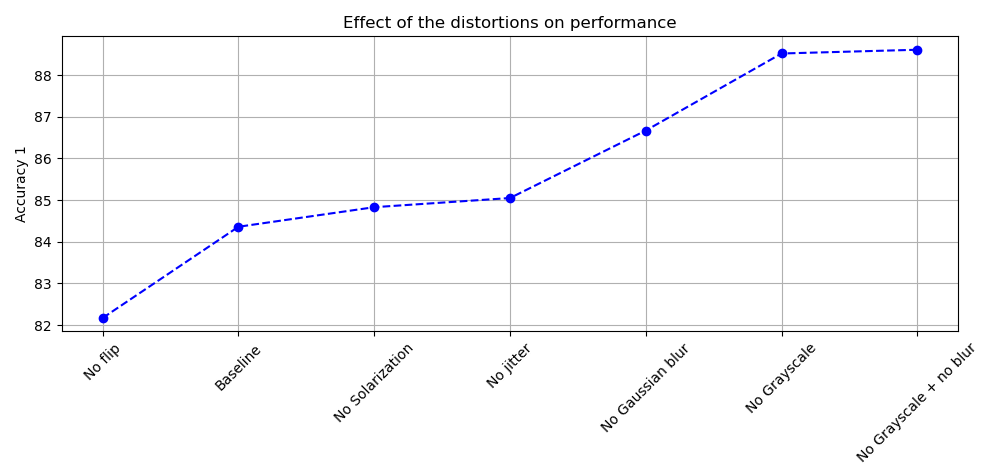}
\end{center}
  \caption{Impact of the different components of the augmentation strategy on the performance. We observe that the baseline, proposed in the original work \cite{BT} does not perform well on our benchmark dataset.}
  \label{fig_transo_ab}
\end{figure}

We conducted a subsequent ablation study to investigate data augmentation strategies' impact further. 
Firstly, we evaluate the default augmentation strategy of Barlow Twins, defined in \cite{BT}. This augmentation strategy includes random crop, horizontal flip, colour jitter, random grayscale, Gaussian blurring and solarization, and normalization. According to the original work, this augmentation strategy is best suited when dealing with natural images. However, when applying this strategy to our dataset, we observe in Figure \ref{fig_transo_ab}, that the baseline performance is the second worst. The only case with worse performance occurs when horizontal flips are removed. Eliminating solarization, colour jitter, Gaussian blur, and grayscale enhances performance, as these augmentations alter the intrinsic colours within each patch. This alteration compromises crucial information regarding specific tissue characteristics carried by the stain's colour.

To adapt this augmentation strategy to pathology, we first propose to study the existing transformation by analyzing the effect of different thresholds on the performance. We conclude that solarization with a high threshold (250) and weak jitter (as supported by \cite{Benchmark}) benefits the training. However, Gaussian blurring, grayscale, or asymmetric crops, such as in DINO \cite{DINO}, do not fit the pathology data. 

\begin{figure*}[htp]
    \centering
    \includegraphics[scale=0.27]{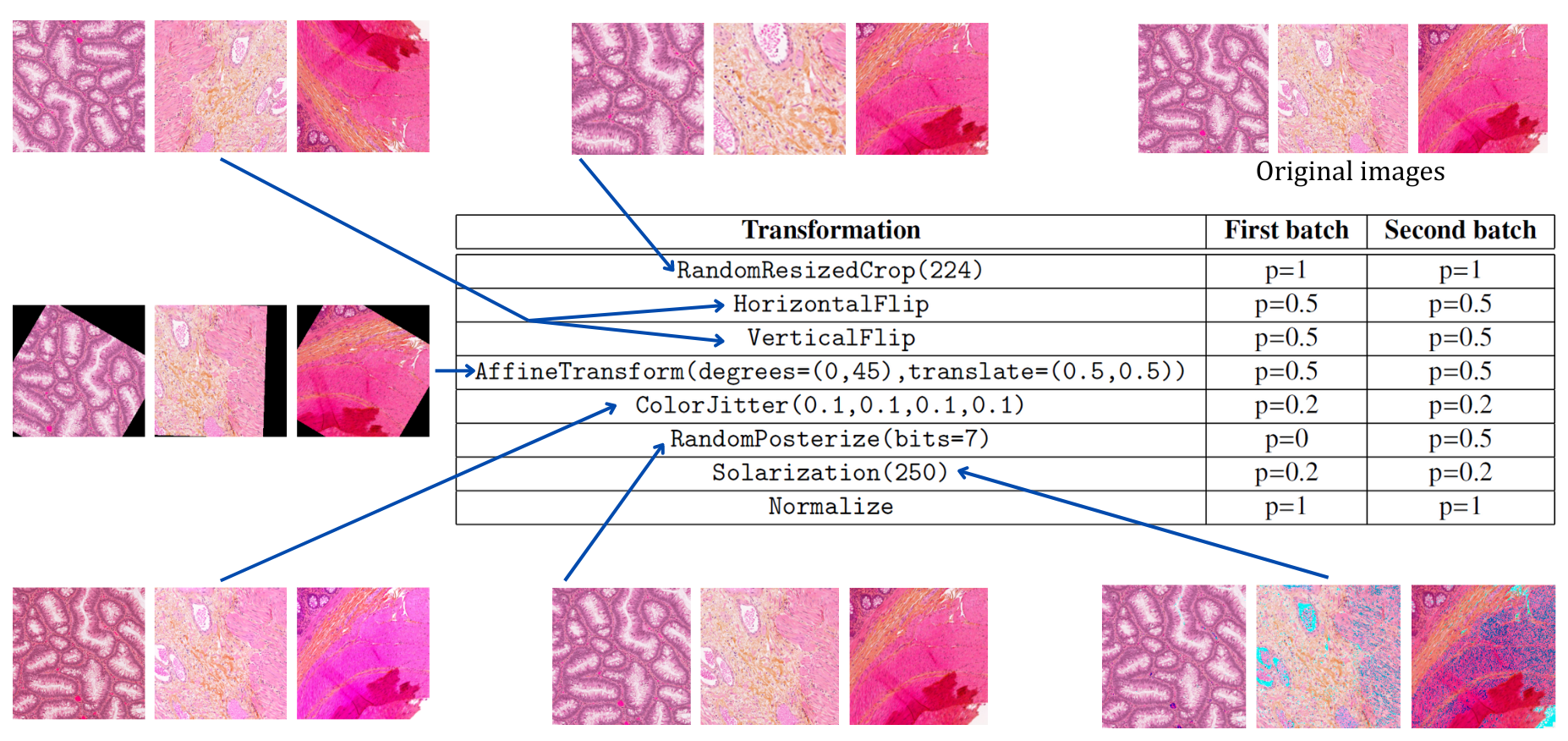}
    \caption{Two sets of new transformations are applied to the input of Barlow Twins to create distorted views. Examples of each transformation applied to different types of tissues are given as examples.}
    \label{fig_table_transfo}
\end{figure*}

Secondly, we propose to study the potential of new transformations in this framework. We conclude that high posterization on one of the batches, vertical flips and affine transformations boost the training. We conduct an in-depth analysis of this ablation study in Section \ref{sec_ab_hp_transfo}. We propose a new augmentation strategy, presented in Figure \ref{fig_table_transfo}.

\subsubsection{Encoder}
Vision Transformers (ViTs) have shown promising results in various computer vision tasks. The Swin Transformer processes images by merging patches in deeper layers \citep{Swin}. Applied to pathology, the Swin Transformers acts as a local-global feature extractor and enhances the learning of both local fine structure and global context \citep{WANG2022102559}. In this last part, and for the first time in CPath, we want to study the potential of the Swin encoder in the Barlow Twins setting. For a fair comparison with ResNet-50 (26 million of trainable parameters), we study Swin-Tiny (28 million of trainable parameters).
\subsection{Evaluation details}
\begin{figure*}[htp]
\begin{center}
  \includegraphics[scale=0.4]{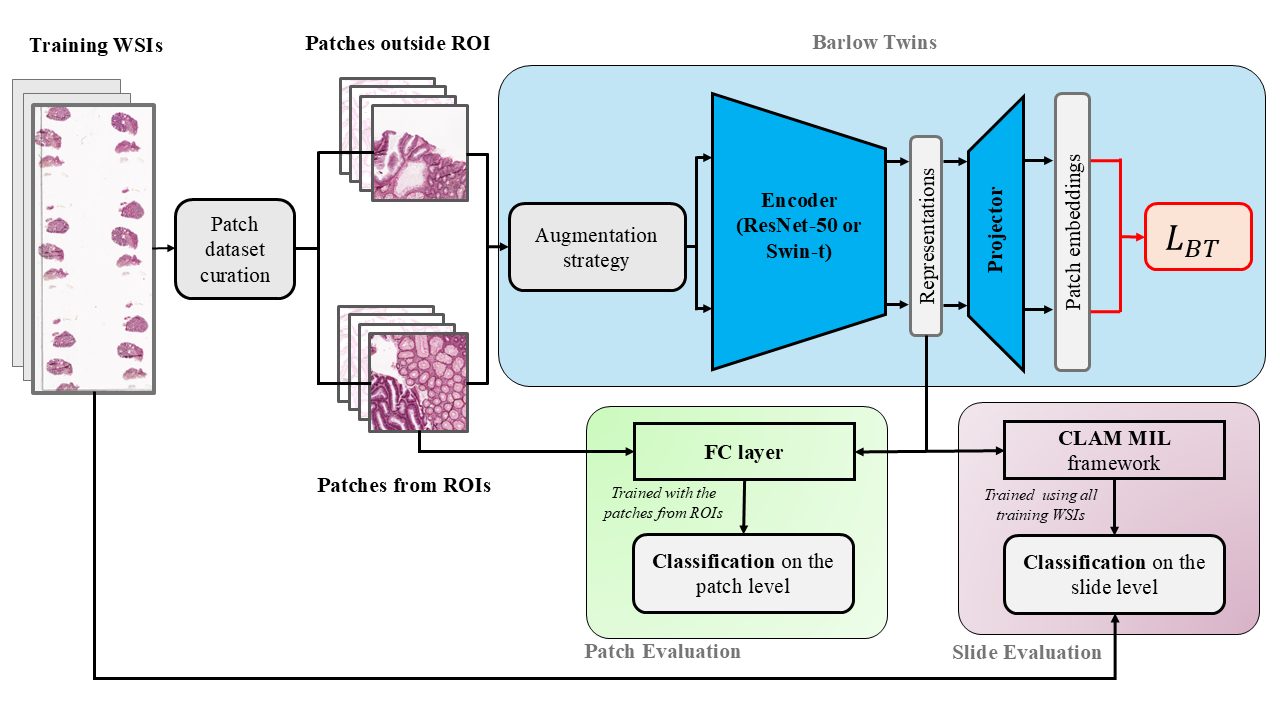}
\end{center}
  \caption{The WSI are tiled into patches from ROI annotations or non-annotated regions. All patches are forwarded into the Barlow Twins framework. The input patches undergo two sets of distortions. The distorted patches are forwarded through an encoder, and their representations are projected into the loss space using a projector. The Barlow Twins loss $L_{BT}$ aims at making the cross-correlation matrix of these two embeddings close to the identity matrix. The learned representations are then evaluated on the patch level by training a Fully Connected Layer on top of the linear encoder and on the slide level using CLAM}
  \label{fig_framework}
\end{figure*}

This work aims to compare the different methods mentionned above: pretraining, augmentation strategy, and encoders. Figure \ref{fig_framework} presents the general framework, from the slide to the patch and slide classification, using the Barlow Twins method.
\subsubsection{Pretraining details}
We train Swin-Tiny and ResNet-50 in a supervised manner, using Normal and ROI patches for 20 epochs, and in a self-supervised manner, using Barlow Twins and all patches (Normal, ROI and non-ROI). More precisely, we train six models, defined in Table \ref{tab_all_models}. For the training in Barlow Twins settings, the LARS optimizer is used with a batch size of 512 with initial learning rates of 0.2 for the weights and 0.0048 for the biases. The projection head is of dimensions 8192. The encoder is trained for 100 epochs. 
\begin{table}[h]
    \centering
    \scriptsize
    \begin{tabular}{|c|c|c||c|c|c|c|}
    \hline
         & $RN50_s$ & $Swin_s$ & \textbf{basicBT} & \textbf{imBT} & \textbf{pathBT} & \textbf{swinBT} \\ \hline \hline
         \textbf{Training} & sup & sup & BT & BT & BT & BT \\ 
         \textbf{Pretraining} & Y & Y & N & Y & Y & Y \\
         \textbf{Augmentation} & N & N & N & N & Y & Y \\
         \textbf{Encoder} & RN50 & Sw-T & RN50 & RN50 & RN50 & Sw-T \\ \hline
         
    \end{tabular}
    \caption{Different trained encoders. $RN50_s,Swin_s$ correspond to ResNet-50 and Swin-Tiny trained in a supervised manner. For the pretraining, Y corresponds to an initialization of the weights with ImageNet weights and N to random initialization. For the augmentation, Y corresponds to the new augmentation introduced in Section \ref{sec_aug} and N corresponds to the default augmentation strategy.}
    \label{tab_all_models}
\end{table}

Additionally, we will compare the performance of these models to those of \textbf{benchBT}, the ResNet-50 encoder pre-trained on more than 36,000,000 WSIs from \cite{Benchmark}.
\subsubsection{Linear evaluation}
To evaluate the quality of the representations learned during the first phase, we perform a linear evaluation. We train a linear classifier on top of a frozen encoder to map the representations to the 5 classes. The top-1 accuracy and the Area Under the Curve (AUC) are the metrics used for comparison. While accuracy is a straightforward metric, it lacks insights into the specific errors made by the model. The AUC provides insight into the trade-off between sensitivity and specificity and can quantify the classifier’s overall performance. Therefore, employing accuracy and AUC provides a more comprehensive understanding of our classifier’s performance. The linear classifier is trained for 100 epochs using the SGD optimizer, the Cosine Annealing scheduler, with an initial learning rate of 0.3. Section \ref{sec_ab_curves} provides more details about this setup. The datasets used for linear evaluation are balanced training sets of 3,500 patches per class and balanced test sets of 300 patches per class, randomly selected.
\subsubsection{Multi Instance evaluation}

To assess the quality of the learned representations for slide-label classification, we employ CLAM method \citep{Clam}. The performance of CLAM will be evaluated using the AUC metric. Moreover, we will compare the generated WSI attention maps with the ROI annotations to assess the enocoders' relevance to the classification task.

\subsection{Ablation Studies}
Due to space limitations, we place the ablation studies in the Appendix. We first introduce the preprocessing steps in Section \ref{sec_ab_data}. Then, in Section \ref{sec_ab_hp_transfo}, we discuss the diverse augmentation and hyperparameters effect. We also present the training curves of the supervised and self-supervised encoders, the linear layers, and the CLAM framework in Section \ref{sec_ab_curves}. We share some additional results, such as confusion matrices and additional heatmaps in \ref{sec_ab_res}.

\section{Experimental Results}
\label{sec_results}
In this section, we first carry out various experiments for colorectal polyps classification at the patch and slide levels. Then, we emphasize the importance of the FoV used for pretraining by evaluating the encoders on diverse FoV. We analyze CLAM heatmaps with respect to the partial annotations of the KGH dataset. To conclude, we reproduce the patch classification on PCam \cite{pcam} benchmark dataset and perform downstream tasking on MHIST \cite{mhist} and NCT-CRC-7k \cite{crc-dataset} leveraging the encoders trained on the KGH dataset. 
\subsection{Patch classification}
\begin{figure*}
    \centering
    \includegraphics[scale=0.45]{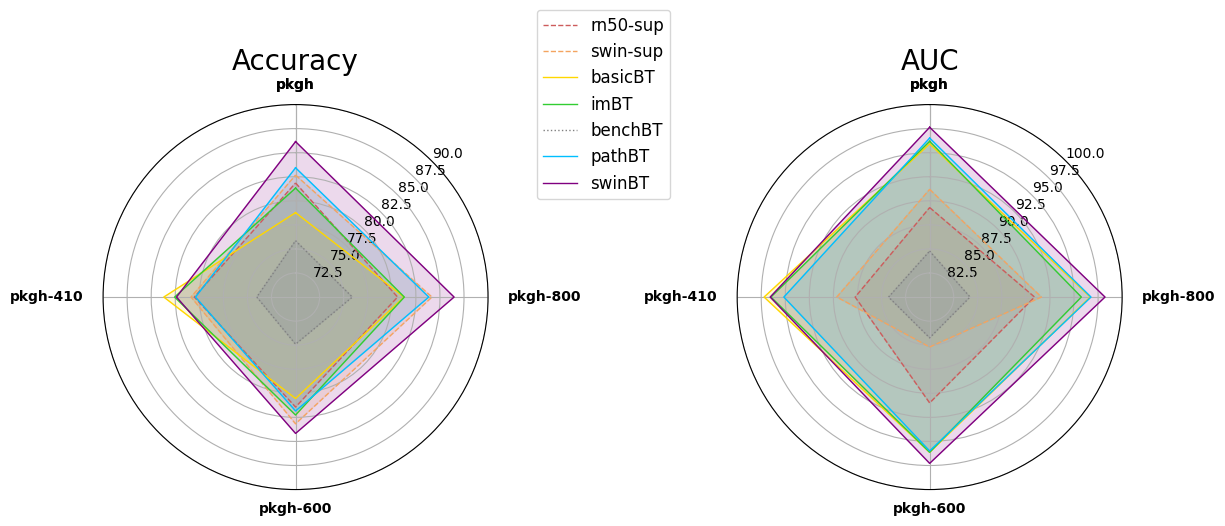}
    \caption{Polar graphs of the Accuracy and AUC of the patch classification for the different models and datasets. The proposed models, in blue and purple, perform relatively well and outperform other models with the same encoder for three out of the four datasets.}
    \label{fig_patch_polar}
\end{figure*}

The classifier is trained two times on the same encoder. Figure \ref{fig_patch_polar} presents the accuracy and AUC for the patch classification. The encoders are evaluated after 10, 50 and 100 epochs of self-supervised training. The epoch of self-supervised training for which the model performs the best is also reported.  We observe that in all cases, the self-supervised methods provide better results than the supervised baseline. Then, we observe that the benchmark weights do not generalize well to our dataset under linear evaluation. Additionally, we can observe that for pkgh and pkgh-800, the two proposed models perform better than the other ones. However, for pkgh-600, pathBT is outperformed by imBT, which is still outperformed by SwinBT. Results for pkgh-410 are surprising: basicBT outperforms imBT, pathBT and swinBT. The ImageNet initialization did not benefit the training for this large dataset. The new data augmentation strategy did not benefit the quality of the extracted features either. The proposed methods perform well on smaller datasets (with less than 500,000 patches) and bigger FoVs (minimum 800 $\mu m$). For large datasets (1.6 million images), ImageNet initialization makes it hard for the model to converge properly. For all models, swinBT outperforms all ResNet-50-based models on the patch level.

\begin{figure}[h]
\begin{center}
  \includegraphics[scale=0.3]{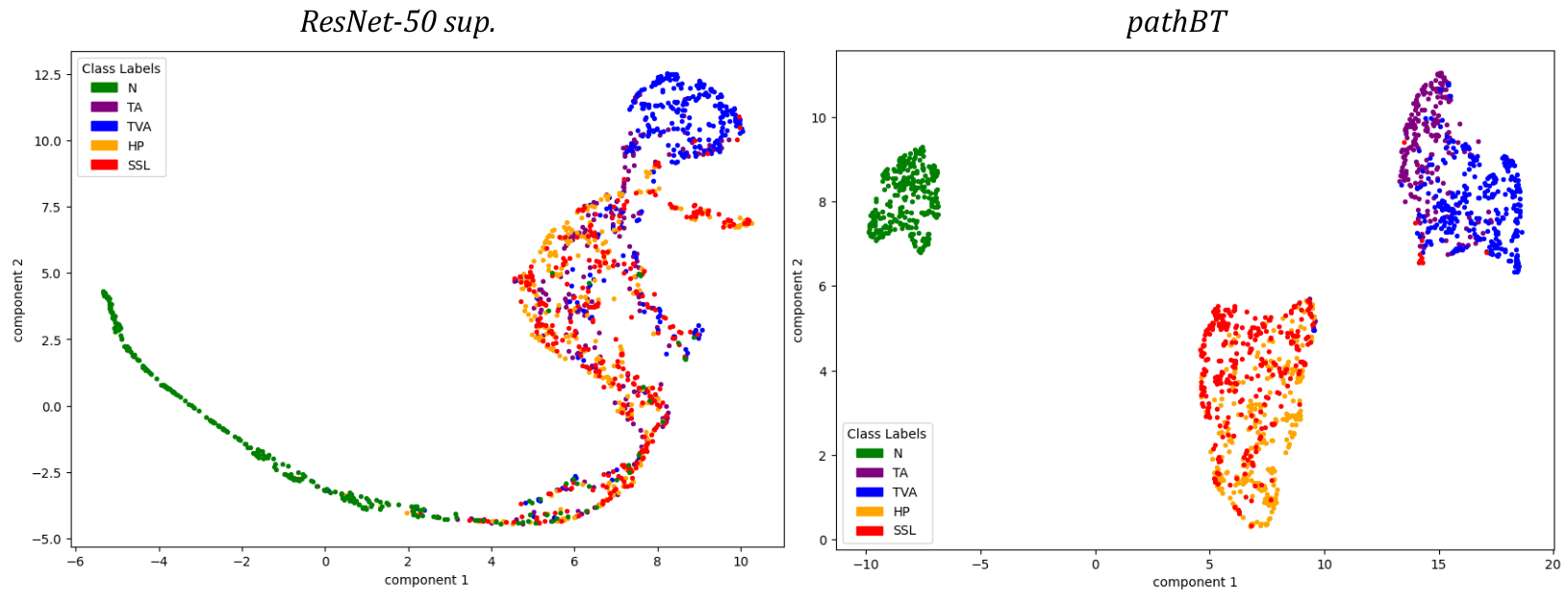}
\end{center}
  \caption{UMAP visualizations for the ResNet-50 encoders trained in a supervised way (on the left) and within pathBT settings (on the right). Here, SSL corresponds to the Sessile Serrated Lesions. We observe that the self-supervised method yields more distinct and separable representations than the supervised method. Furthermore, the three groups identified by pathBT, Normal, HP/SSLe and TA/TVA, align with the diagnostic ambiguities observed in the data.}
  \label{fig_umaps}
\end{figure}

Figure \ref{fig_umaps} displays the UMAP visualizations of the classified features for the ResNet-50 trained supervised and in pathBT settings. The UMAP for ResNet-50 supervised shows that the Normal class is quite separated from the rest, but the four other classes are still mixed. However, the UMAP for the pathBT model shows three distinct groups: Normal, HP/SSLe and TA/TVA, meaning that the model struggles to separate HP from SSLe and TA from TVA. This visualization suggests that HP and SSLe share a similar feature. The same conclusion can be drawn for TA and TVA. These results can be explained by the fact that TVA is TA with more than 20\% of villous architecture. Therefore, TA structures can still have a bit of villous architecture. Additionally, SSLe is more common in the right colon, and HP in the left \citep{pickhardt2018natural}. By default, a growth in the right colon will be considered SSLe until proven otherwise. The fact that the model is able to catch this ambiguity shows that it is learning meaningful features.

\subsection{Slide classification}
\begin{figure*}
    \centering
    \includegraphics[scale=0.35]{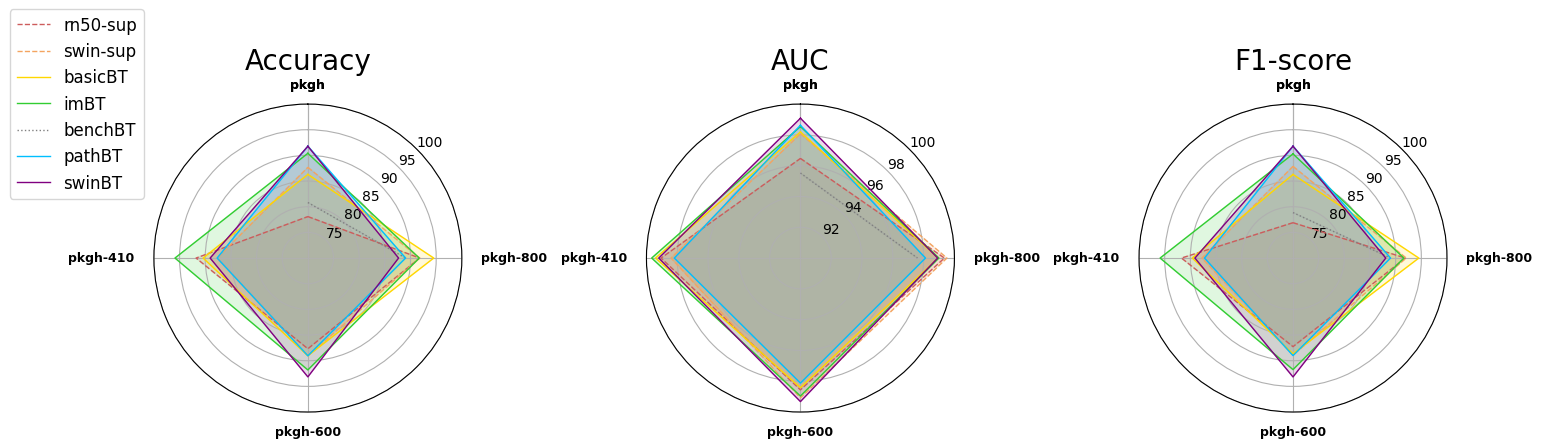}
    \caption{Polar graphs of the Accuracy and AUC of the slide classification for the different models and datasets. The results do not follow the same trends as in Figure \ref{fig_patch_polar}. However, we remark that swinBT performs relatively well. Slide evaluation does not rely on the same features as the patch classification. }
    \label{fig_slide_polar}
\end{figure*}

If patch results can help guide pathologists at the microscopic level, performing a slide evaluation
can help pathologists by sorting the slides by order of importance. In this section, the MIL framework, CLAM \cite{Clam}, is trained once on the same encoder. Figure \ref{fig_slide_polar} presents the slide classification's accuracy, AUC and F1 score. Here, the results are more nuanced. The benchmark weights do not provide a good enough representation of the data and the MIL model fails to converge for pkgh-600 and pkgh-410. If the supervised baselines do not perform greatly overall, the supervised ResNet-50 reaches the second-best AUC on pkgh-600 and the second-best accuracy and F1-score on pkgh-410, emphasizing the possibility of training models on a subset of the training set. However, the self-supervised models generally perform better than the supervised models. For pkgh, the two proposed methods perform better on all metrics. SwinBT performs better than pathBT in terms of AUC metrics and always provides a very satisfying AUC above 0.9891. For pkgh-800, it seems that basicBT surpasses all models for accuracy and F1-score. For pkgh-600, pathBT is outperformed by imBT on all metrics, but swinBT provides the best results. The best classification overall is obtained by imBT encoder on pkgh-410 dataset. 
\subsection{Pretraining on Different Fields of View}
In this section, we aim to utilize encoders pretrained on one FoV, to classify patches or slides at a different FoV. For linear evaluation, we freeze the pathBT encoder trained on pkgh, pkgh-600 and pkgh-800 and we train a linear layer on a dataset with a different FoV.
\begin{table}[htp]
    \centering
    \footnotesize
    \begin{tabular}{|c|c||c|c|c|}
    \hline
         && \textbf{pkgh} & \textbf{pkgh-800} & \textbf{pkgh-600}  \\ \hline \hline
        Trained&\textbf{pkgh}& \textcolor{gray}{83.43 - \textit{0.9652}} & 77.5 - \textit{0.9399}& 75.3 - \textit{0.9472} \\
        on &\textbf{pkgh-800} & 79.76 - \textit{0.9591} & \textcolor{gray}{83.84 - \textit{0.9676}} & 82.34 - \textit{0.9608} \\
         & \textbf{pkgh-600} & 78.14 - \textit{0.9509}& 80.17 - \textit{0.9589}& \textcolor{gray}{81.84 - \textit{0.9599}} \\ \hline
    \end{tabular}
    \caption{Accuracy and \textit{AUC} (acc - \textit{AUC}) for patch classification of the different pathBT encoders evaluated on different FoV}
    \label{tab_t_pathbt}
\end{table}
\begin{table}[htp]
    \centering
    \footnotesize
    \begin{tabular}{|c|c||c|c|c|}
    \hline
         && \textbf{pkgh} & \textbf{pkgh-800} & \textbf{pkgh-600} \\ \hline \hline
        Trained &\textbf{pkgh}& \textcolor{gray}{86.15 - \textit{0.9764}} & 83.33 - \textit{0.973}& 81.33 - \textit{0.967} \\
        on & \textbf{pkgh-800} & 85.05 - \textit{0.9757} & \textcolor{gray}{86.47 - \textit{0.9822}} & 82.8 - \textit{0.9666}  \\
         &\textbf{pkgh-600} & 83.33 - \textit{0.9688} & 85.73 - \textit{0.9754}& \textcolor{gray}{84.17 - \textit{0.9728}} \\
       \hline
    \end{tabular}
    \caption{Accuracy and \textit{AUC} (acc - \textit{AUC}) for patch classification of the different swinBT encoders evaluated on different FoV}
    \label{tab_t_swinbt}
\end{table}

Table \ref{tab_t_pathbt} shows the transferability results of pathBT models. Each model performs better on the model they were trained on using Barlow Twins rather than on the model they are evaluated on. However, pathBT trained on pkgh-800 performs better on pkgh-600 than pathBT trained on pkgh-600. It is possible that models trained on pkgh-600 or smaller struggled to generalize to new data.

Table \ref{tab_t_swinbt} shows the transferability results of swinBT models. The overall results are better than those with pathBT, highlighting the ability of the Swin Transformer to generalize to unseen data and its pertinence for pathology data.

We also observe that the models generalize better to the datasets whose FoV are close to the FoV used during pretraining, highlighting the fact that the features learned by the models are linked to the FoV and the visible structures in the patches. 

Secondly, on the slide level, we train the CLAM framework \cite{Clam} on top of the frozen encoders. Table \ref{tab_cross_wsi_swin} shows that a closer FoV to the one used during pretraining will provide better results than a much different FoV. This is highlighting again the fact that the features learned by the models are linked to the FoV. 

\begin{table}[htp]
    \centering
    \footnotesize
    \begin{tabular}{|c|c|c|c|c|}
    \hline
        & & \textbf{pkgh} & \textbf{pkgh-800} & \textbf{pkgh-600} \\ \hline \hline
        Trained & \textbf{pkgh} & \textcolor{gray}{91.78 - \textit{0.9908}} & 91.78 - \textit{0.9858} & 90.41 - \textit{0.9912} \\
        on & \textbf{pkgh-800} & 91.78 - \textit{0.9805} & \textcolor{gray}{87.67 - \textit{0.9891}} & 89.04 - \textit{0.9892} \\
        & \textbf{pkgh-600} & 84.93 - \textit{0.9726} & 89.04 - \textit{0.9892} & \textcolor{gray}{93.16 - \textit{0.9932}} \\ \hline
    \end{tabular}
    \caption{Accuracy and \textit{AUC} (acc - \textit{AUC}) for the slide classification of the different swinBT encoders evaluated on different FoV using CLAM framework}
    \label{tab_cross_wsi_swin}
\end{table}

These results also emphasize that generalizable models should be able to adapt to different FoV and, therefore, be trained on different FoV. This observation aligns with the observation made by \cite{Benchmark}.

\subsection{CLAM heatmaps and ROI Correlation}
\label{sec_clam_results}
\begin{figure}[h]
\begin{center}
  \includegraphics[scale=0.4]{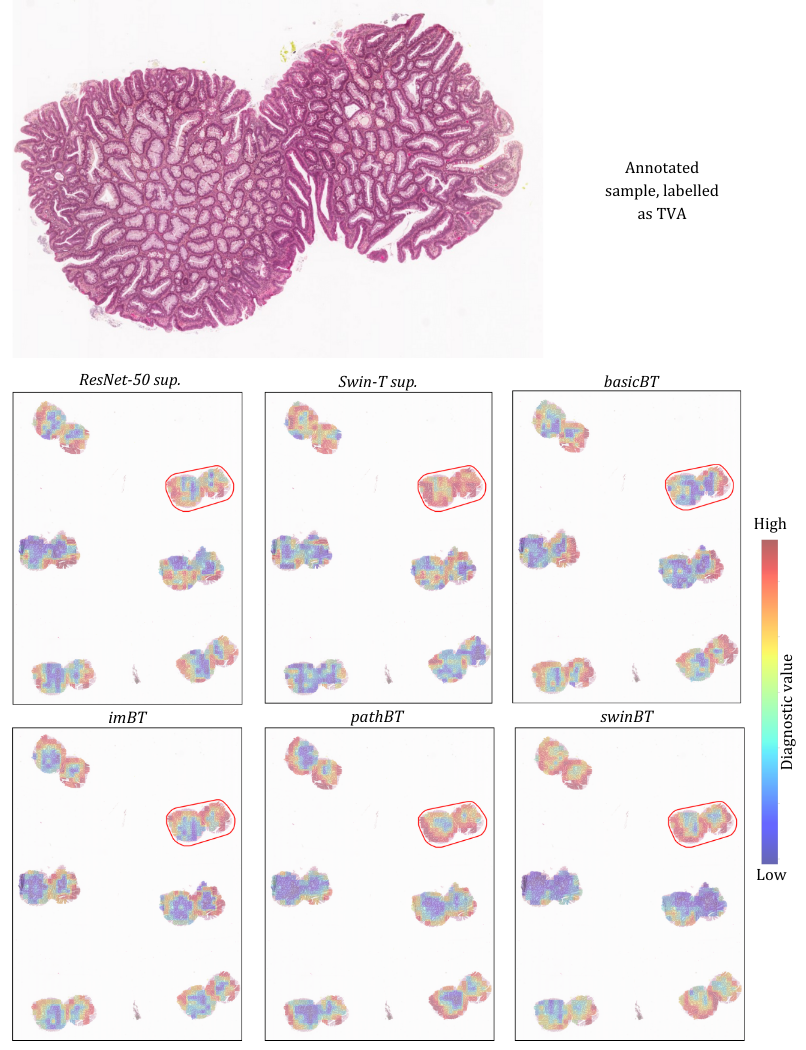}
\end{center}
  \caption{A TVA WSI annotated by the different models. We observe that the different models highlight different regions of high interest. The sample above is a zoomed-in view of the annotated sample. }
  \label{fig_slides_rois_2}
\end{figure}

A strength of the CLAM framework is to determine areas with high diagnosis value. However,
this asset is not taken into consideration here. As the analysis of the decision of the CLAM
framework carries meaningful explanations, Figure \ref{fig_slides_rois_2} shows the different CLAM heatmaps of a TVA slide after the MIL framework on six different encoders. First, all heatmaps are different despite correlating with the pathologist annotation (red line). After clarification with our expert pathologist, we found that these tissue samples all comprise villous features at the bottom border of all tissue samples (as shown by the pathBT model, for example). The center of the samples is normal tissue. Both supervised encoders highlight patches within the ROI. However, as mentionned in section \ref{sec_data}, ROI annotations should be transferrable to the other tissue samples below the top right one.  For the two supervised models, the heatmap within the ROI annotation does not map to the heatmaps from the samples underneath. For the Barlow Twins encoders, mainly for imBT, pathBT and swinBT, the local portions of the heatmaps corresponding to the ROI are reproducible to other tissue samples in the slide. This result shows that the self-supervised encoders caught the general organization of the slide, while the supervised ones, only fed with patches from the non-ROI, tend to generalize to the whole slide.
To complete this analysis, Figure \ref{fig_patches} shows the six patches with the highest diagnosis values. We observe that patches of the highest diagnosis values for Resnet-50 supervised, basicBT and imBT present artifacts, such as tissue fold and torn tissue. However, the supervised swin-T, pathBT, and swinBT do not focus on these patches, and the final diagnosis is made using relevant tissue patches. Moreover, pathBT and swinBT share three patches of high diagnosis values in a different order. This observation underscores the effectiveness of the proposed augmentation strategy in bridging the differences between encoders. 

\begin{figure}[htp]
\begin{center}
  \includegraphics[scale=0.28]{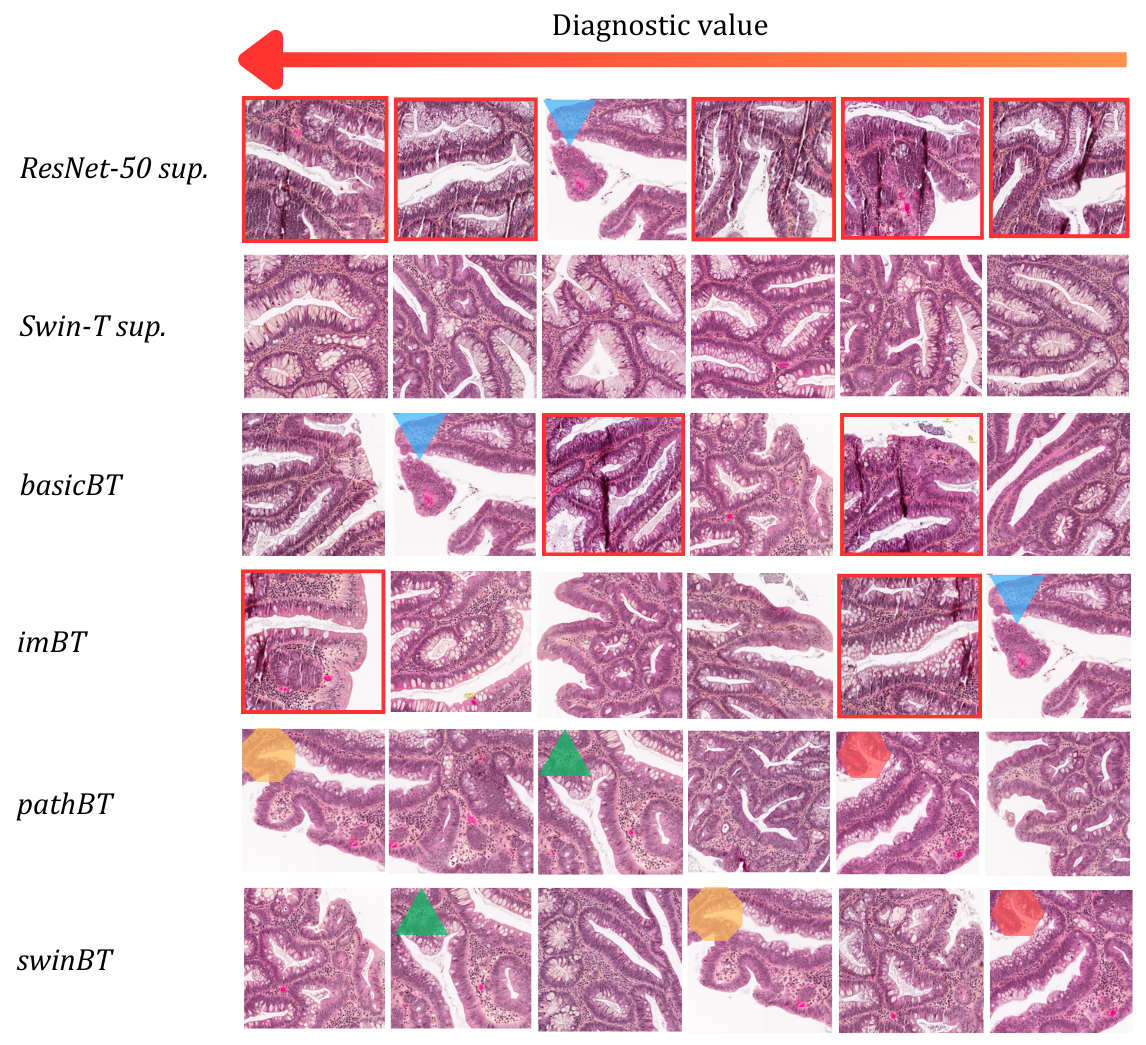}
\end{center}
  \caption{Six patches with the highest diagnostic values for the six models across the slide showing TVA, presented in Figure \ref{fig_slides_rois_2}. The patches with the highest diagnostic value are on the left. Coloured shapes highlight that the same patches were found to be of high diagnosis value across different encoders. The red contour highlights the noisy patches.}
  \label{fig_patches_fov}
\end{figure}
\subsection{Patch classification on PCam}
As the proposed augmentation strategy was studied and established on a subset of the KGH dataset, it seems fair to reproduce the experiments described in Table \ref{tab_all_models} on another dataset. PCam is a well-known challenging benchmark for binary class breast cancer type classification and is, therefore, a robust dataset to evaluate our method. To reproduce similar settings, where the encoder is trained on all available data while the linear evaluation is performed on limited annotated data, the encoder is trained on the PCam training dataset while the linear layer is trained on a random 5\% of the PCam training set. The encoder was trained once for 100 epochs with a batch size of 256 to alleviate the need for computational resources, and the linear layer was trained for 100 epochs on the same frozen encoder with two different folds of 5\% of the training set. The model is then evaluated on the PCam test set. Table \ref{tab_pcam} presents the accuracy and AUC of the different methods. 
\begin{table}[h]
    \small
    \centering
    \begin{tabular}{|c|c|c|c|c|}
    \hline 
         & \textbf{basicBT} & \textbf{imBT} & \textbf{pathBT} & \textbf{swinBT} \\ \hline \hline
        \textbf{ACC} & 69.2 & 69.84 &  70.5 & 79.63 \\
        \textbf{AUC} & 0.692 & 0.6984 & 0.705 & 0.7963 \\ \hline
    \end{tabular}
    \caption{Accuracies and AUC of PCam test set after linear evaluation on the encoders trained in the settings mentionned in Table \ref{tab_all_models} with a batch size of 256. The training set was used to train the encoder and 5\% of this training set was used for training the linear layer.}
    \label{tab_pcam}
\end{table}

We observe that imBT performs better than basicBT (+0.6\%). Therefore, it seems that the weight initialization still matters. Secondly, PathBT performs slightly better (+0.6\% in accuracy and AUC) than basicBT and imBT; therefore, the augmentation strategy developed for the KGH dataset can still be applied to this dataset. Additionally, SwinBT is still outperforming all other methods (+9\%), highlighting the potential of Swin transformers for pathology data.
\subsection{Downstream tasking on MHIST and CRC}
In this section, the capacity of our model to generalize to other colon datasets is evaluated. We train a linear layer on top of the frozen encoders to map the representations to the classes for two datasets: MHIST \cite{mhist} and NCT-CRC-7k \cite{crc-dataset}. 

\textbf{MHIST.} MHIST dataset is designed to classify colorectal polyps into two histological categories: Hyperplastic Polyps and Sessile Serrated Adenomas. It comprises 3,152 patches which are H\&E stained, and it was chosen because of the similarity of these two classes to those of our KGH dataset. The original split train/test is used to train and evaluate the linear layer. Table \ref{tab_mhist} presents the accuracy and AUC of the diverse models pretrained on three different FoV of the KGH dataset using four different methods presented in Table \ref{tab_all_models}. Additionally, the linear evaluation of benchBT on MHIST provides an accuracy of 80.76 and an AUC of 0.7943. For MHIST dataset, basicBT and imBT outperform the benchmark weights and the two proposed methods. If basicBT provides the best accuracy (for the model pre-trained on pkgh-600), imBT provides the best AUC (for the model pre-trained on pkgh-800). If the accuracy for imBT remains constant throughout the three datasets, it seems the representations given by basicBT, pathBT and swinBT are highly dependent on the pretraining dataset. These results show that our proposed methods do not generalize well to this downstream task. However, the default transformations generalize better. The weak colour jitter could be a reason for this lack of generalizability, as the model will have a lower ability to generalize to other stains. These results also highlight the potential of the KGH dataset to be used as a pretraining dataset for diverse downstream tasks, as imBT and basicBT outperform the benchmark model, trained on a massive cohort of data (36k WSIs).
\begin{table}[h]
    \small
    \centering
    \begin{tabular}{|c|c|c|c|}
    \hline
         & \textbf{pkgh} & \textbf{pkgh-800} & \textbf{pkgh-600} \\ \hline \hline
         \textbf{basicBT} & \textbf{82.5 - \textit{0.8061}} & \textbf{83.32 - \textit{0.8194}} & \textbf{84.95 - \textit{0.7971}} \\ 
         \textbf{imBT} & 82.5 - \textit{0.8033} & 82.6 - \textit{0.8264} & 82.5 - \textit{0.8217} \\
         \textbf{pathBT} & 76.05 - \textit{0.7563} & 74.62 - \textit{0.7392} & 75.54 - \textit{0.7361} \\
         \textbf{swinBT} & 79.12 - \textit{0.7634} & 76.56 - \textit{0.7408} & 77.79 - \textit{0.7628} \\ \hline
    \end{tabular}
    \caption{Accuracy - \textit{AUC} for downstream tasking on MHIST dataset using the encoders pretrained on three FoV of the KGH dataset using different methods}
    \label{tab_mhist}
\end{table}

\textbf{NCT-CRC.} The NCT-CRC-7k dataset is the validation set of the NCT-CRC-100k dataset and is made of 7,180 H\&E patches containing human colorectal cancer and normal tissue patches. This dataset comprises nine classes (adipose, background, debris, lymphocytes, mucus, smooth muscle, normal colon mucosa, cancer-associated stroma and colorectal adenocarcinoma) and was chosen because this large repository presents tasks similar to the KGH dataet. 50\% of this dataset was used for training the linear layer and 50\% for evaluating the model. Table \ref{tab_crc} presents the accuracy and AUC of the diverse models pretrained on three different FoV of the KGH dataset using four different methods presented in Table \ref{tab_all_models}. Additionally, the linear evaluation of benchBT on the test split of NCT-CRC-7k provides an accuracy of 96.6 and an AUC of 0.9985. The results from Table \ref{tab_crc} are inconsistent with those of MHIST in the precedent section. Here, we observe that swinBT outperforms all other models for all pretraining datasets regarding accuracy. Representations learnt by pathBT are the ones giving the top-2 best AUC across all pre-trained datasets. Here, all four models perform better than the benchmark model for accuracy but not for the AUC (third best). In this case, we show that the two proposed methods align better with the CRC dataset and that the KGH dataset has a high potential for model pretraining for colorectal screening.

\begin{table}[h]
    \centering
    \small
    \begin{tabular}{|c|c|c|c|}
        \hline
         & \textbf{pkgh} & \textbf{pkgh-800} & \textbf{pkgh-600} \\ \hline \hline
         \textbf{basicBT} & 97.99 - \textit{0.9963} & 98.52 - \textit{0.9986} & 99.08 - \textit{0.9995} \\
         \textbf{imBT} & 98.08 - \textit{0.9937} & 98.69 - \textit{0.9966} & 98.8 - \textit{0.9971} \\
         \textbf{pathBT} & 98.47 - \textit{0.9971} & 98.5 - \textit{0.9978} & 98.83 - \textit{0.9986} \\
         \textbf{swinBT} & \textbf{99.11 - \textit{0.998}} &\textbf{ 99.03 - \textit{0.9984}} & \textbf{99.19 - \textit{0.9984}} \\ \hline  
    \end{tabular}
    \caption{Accuracy - \textit{AUC} for downstream tasking on NCT-CRC-7k dataset using the encoders pretrained on three FoV of the KGH dataset using different methods}
    \label{tab_crc}
\end{table}

The evaluation of the encoders pretrained on KGH on these two colorectal polyps datasets has shown that the KGH dataset has a high potential to be used as a dataset for model pretraining for colorectal polyp screening.
\section{Concluding remarks and Future Work}
\label{sec_ccl}
Feature embedding from colorectal polyps for colorectal cancer (CRC) screening relies on efficient and effective representation of Whole Slide Images (WSIs). In this paper, we propose an enhanced Barlow Twins framework for CRC screening by adapting the augmentation strategy to the pathology downstream task and leveraging the Swin Transformer (1). We demonstrate that the proposed models deliver robust performance at both patch and slide levels and generate relevant representations of the patches, facilitating meaningful and accurate diagnoses for WSIs. We also highlighted that supervised methods provide sub-optimal results on our weakly annotated dataset (2). Furthermore, we showed that, for WSI classification, a small Field of View (FoV) of 410 $\mu m$ showed high AUCs above 0.99 across the four datasets, while a larger FoV results in less effective WSI classification (3). We also established that the self-supervised methods learn more realistic, robust and meaningful features than the supervised models (4). We demonstrated that encoders trained on the KGH dataset can be used for model pretraining for colorectal polyp classification (5), and we release the weights. The proposed method has the potential to be applied to a wide range of other pathology-related tasks. We believe that adapting frameworks developed for natural images to pathology data could result in improved performance.

These results motivate future research directions. Firstly, SSL methods need to be tailored to pathology downstream tasks. Investigating the augmentation strategy and the different encoders in DINO \cite{DINO} or SwAV \cite{caron2020unsupervised} could pave the way for other enhanced models in CPath. Secondly, the data augmentation study was done on the patch level, and we observed that the results between the patch and slide levels were not necessarily consistent. Therefore, future works should investigate the best augmentation strategy for slide-level classification. We believe that further exploration of these topics will yield considerable improvements for pathology-specific SSL.

\section*{Acknowledgments}
This work is funded by the Fonds de recherche du Québec - Nature et technologies (FRQNT). In addition, the data collected for this study is supported by Huron Digital Pathology and Ontario Molecular Pathology Research Network (OMPRN) funding grant. We also extend our gratitude to the students at Atlas Analytics Lab, in particular to Damien Martins Gomes and Ali Nasiri-Sarvi, for their fruitful discussions. In addition, we thank Resources for Research Groups(RRG)--Digital Research Alliance Canada (DRAC), Computer Science Cluster at Concordia University, and Huron Digital Pathology for providing essential computational resources. 
\section*{Data availability}
The dataset for this study is offered to editors and peer reviewers at the time of submission for the purposes of evaluating the manuscript upon request. The data is not publicly available due to privacy concerns and Intellectual Property matters related to Huron Digital Pathology.

\appendix

\section*{Supplementary Material}
\label{sec_supp}
\section{More details about the KGH dataset}
\label{sec_ab_data}
\subsection{Colorectal Polyps of the KGH dataset}
\begin{figure*}
\begin{center}
\includegraphics[scale=0.27]{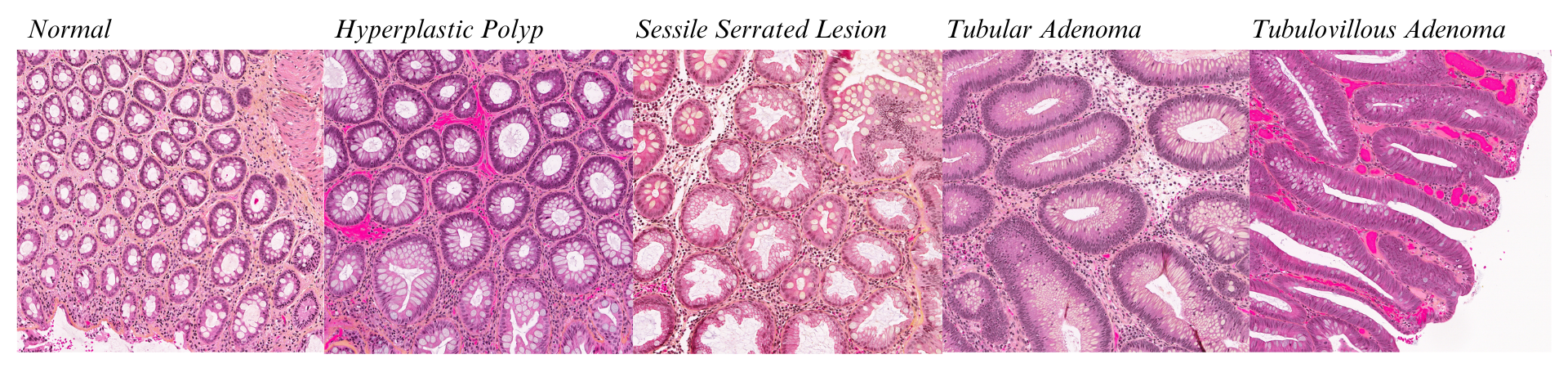}  
\end{center}
\caption{Normal and pathological tissue from our dataset. We can recognize some serrated structures (HP and SSLe) and some tubular structures (TA and TVA)}
\label{fig_polyps_kgh}
\end{figure*}

The focus of the research presented in this thesis primarily revolves around four distinct types of polyps, presented in Figure \ref{fig_polyps_kgh}:
\begin{itemize}
    \item \textbf{Hyperplastic Polyps (HP)}: These noncancerous growths carry a low risk of malignant transformation and are characterized by an overgrowth of normal cells in the mucosal lining of the colon or distal colon; \citep{9,10}
    \item \textbf{Sessile Serrated Lesions (SSLe)}: SSLes, a subtype of serrated polyps, constitute at least 20\% of all serrated polyps. They are considered precancerous and are characterized by a flat or slightly elevated shape, predominantly found in the cecum and ascending colon. Distinguishing SSLes from hyperplastic polyps may pose challenges, although pathologists can recognize certain distortions; \citep{7,8}
    \item \textbf{Tubular Adenomas (TA)}: These are usually small and benign polyps, prevalent in more than 80\% of cases. While they are considered precancerous, less than 10\% of them have the potential to progress into cancer; \cite{4,5,6}
    \item \textbf{Tubulovillous Adenomas (TVA)}: TVAs are a subtype of colonic adenomas exhibiting a combination of tubular and villous features. Considered precancerous, they have the potential to transform into malignant structures. \citep{4,11,12}
\end{itemize}

\subsection{Lightweight CNN for artefact detection}
\begin{figure}[htp]  
\begin{center}
        \includegraphics[scale = 0.3]{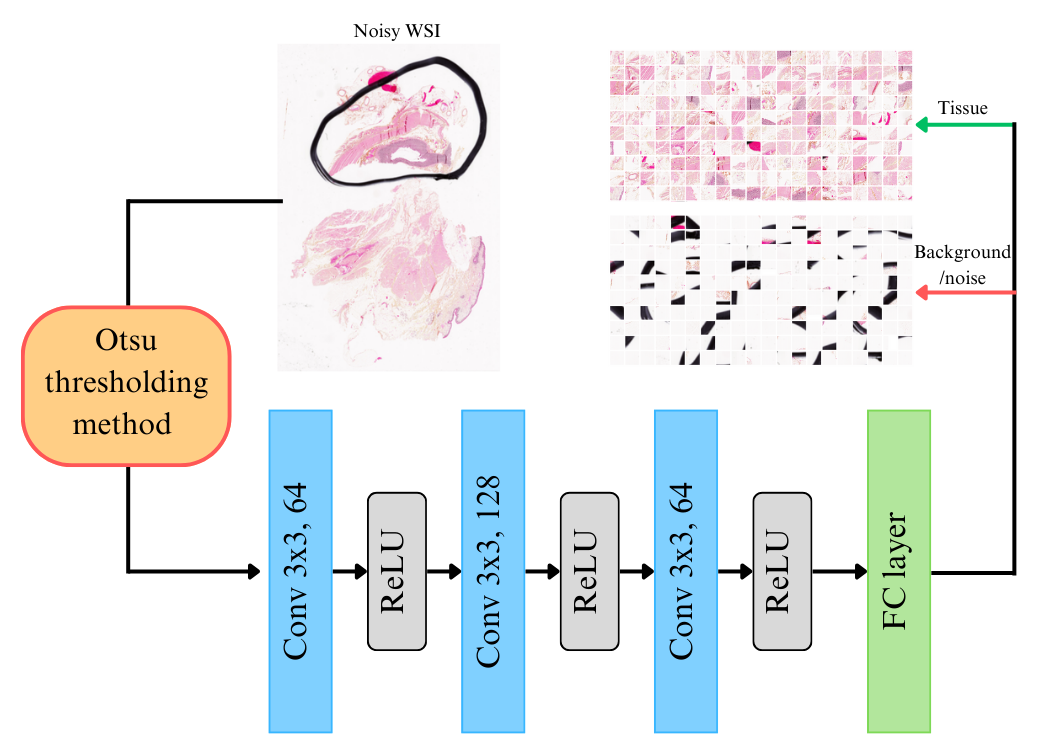}
\end{center}
    \caption{Lightweight CNN for background detection and noise removal. First, patches are extracted from the WSI using the Otsu thresholding concept, integrated into TIAToolbox. Then, the patches labelled as tissue by this first filter are forwarded to a lightweight CNN of three convolutional layers with ReLU activation, which labels the patches as tissue or background/noise.}
    \label{fig_custom_model}
\end{figure}

While state-of-the-art tools, such as Clam \cite{Clam} or TIAToolbox \cite{Pocock2022}, are available for extracting patches from Whole Slide Images for Deep Learning computation, they fail to detect the noise and artifacts present in our dataset, such as glass artifacts and marker annotations. Therefore, we propose a lightweight Convolutional Neural Network (CNN) model that acts as a second filter for the patches extracted by the TIAToolbox Otsu method \cite{otsu}. Figure \ref{fig_custom_model} shows the lightweight CNN. It is made of three convolutional layers with ReLU activation and a Fully Connected layer to map the representations to the two classes (tissue and background/noise). This CNN comprises 469,442 trainable parameters and was trained on a custom dataset made of 10,000 patches from each class, manually extracted at different Fields of View (FoV). It reaches a test accuracy of 98.6\% and is relatively fast, as predicting one patch takes 0.9 ms. The code for this patch extraction tool is available at \href{https://github.com/CassNot/pKGH}{https://github.com/CassNot/pKGH}.

\subsection{Patch extraction}
\begin{figure*}[htp]
\begin{center}
  \includegraphics[scale=0.3]{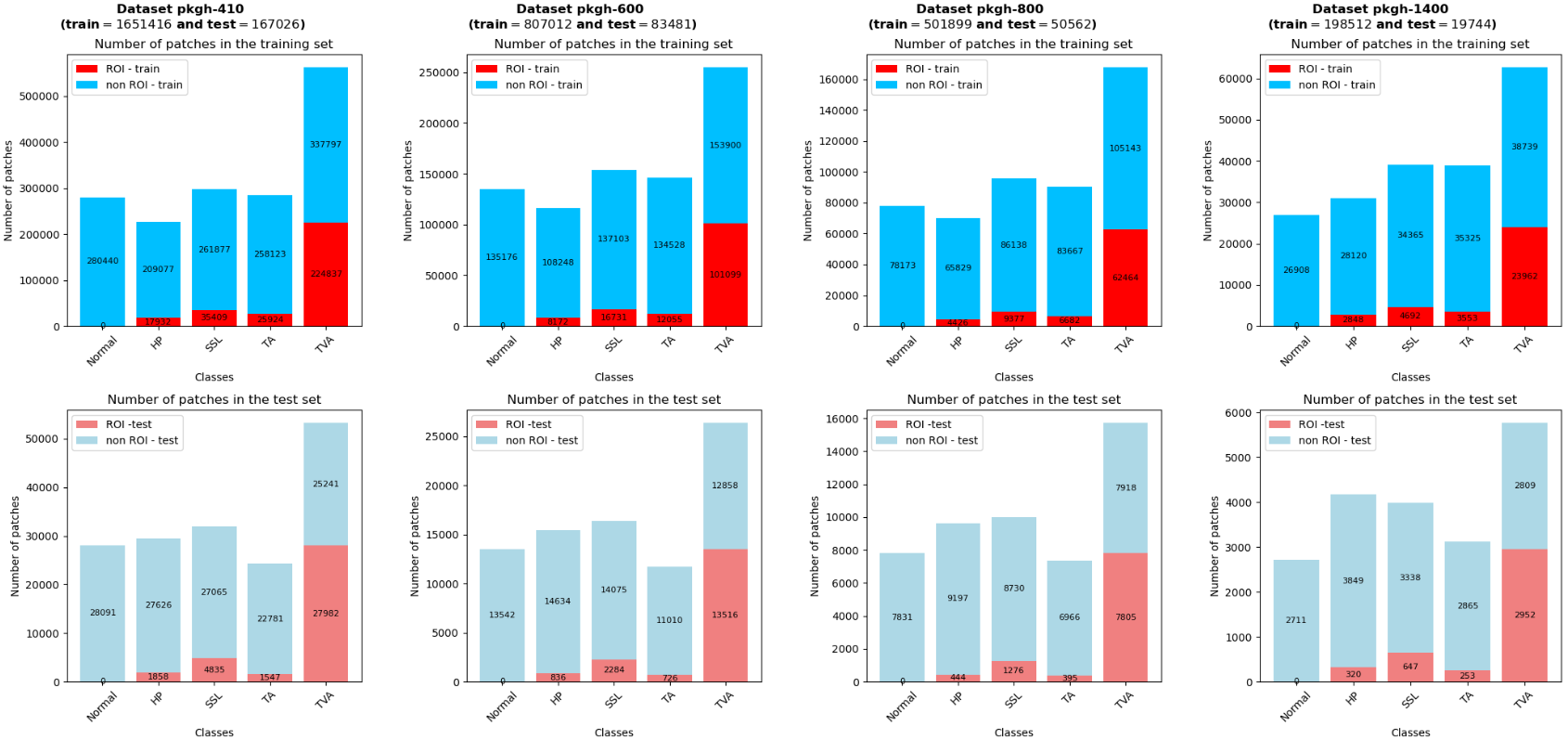}
\end{center}
  \caption{Number of extracted patches within and outside ROI, for the training and test data}
  \label{fig_nb_patches}
\end{figure*}

Using this model, patches from the KGH dataset mentionned in Section \ref{sec_data} have been extracted with four different FoVs: 410 $\mu m$, 600 $\mu m$, 800 $\mu m$ and 1400 $\mu m$. Figure \ref{fig_nb_patches} presents the patches extracted across all train and test slides and shows the counts of patches within and outside of the ROI annotations for each of the five classes. Figure \ref{fig_patches_fov} shows patches from each pathology at the four different FoVs. We can see that a larger FoV shows the general organization of the polyp while a smaller FoV focuses on the specific regions, such as some crypts. 

\begin{figure}[htp]  
\begin{center}
        \includegraphics[scale = 0.3]{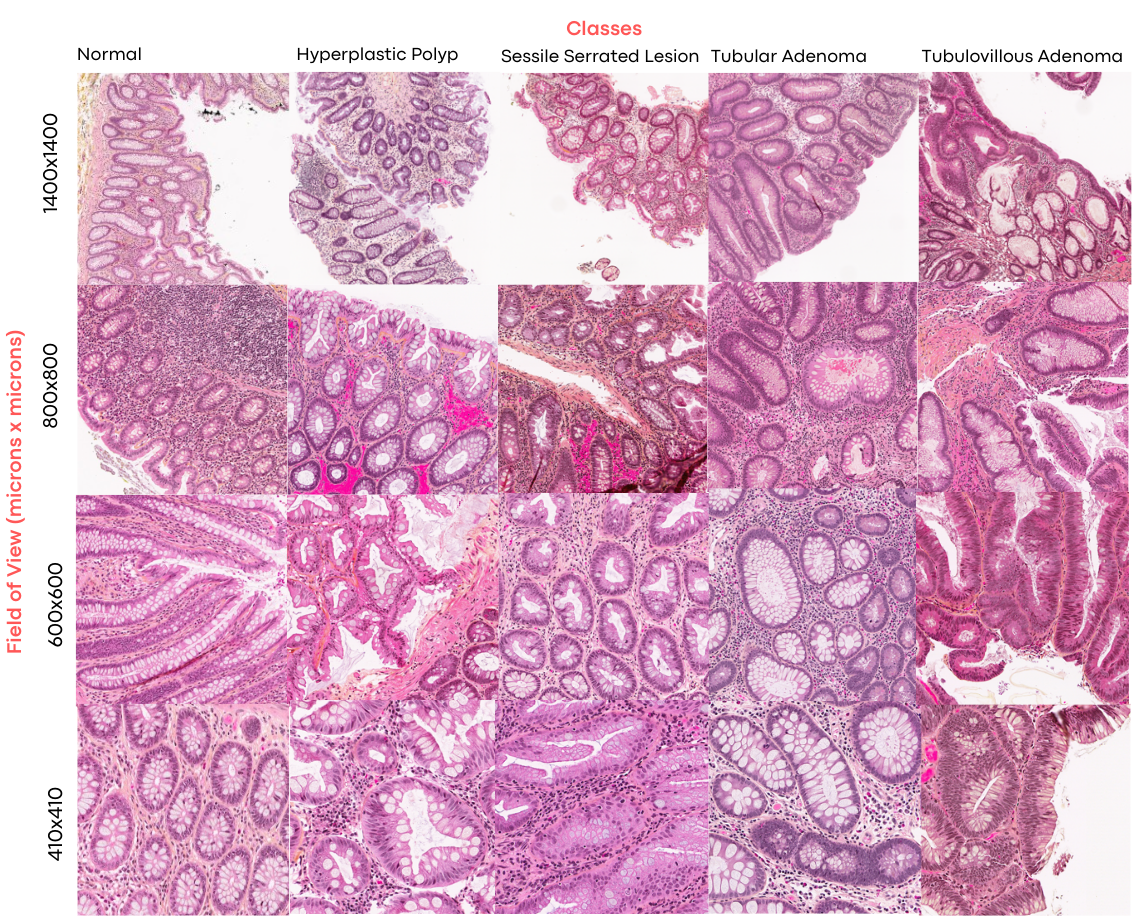}
\end{center}
    \caption{Five classes of the KGH dataset under different FoV in $\mu m \times \mu m$}
    \label{fig_patches}
\end{figure}

\section{Ablation studies for PathBT}
\label{sec_ab_hp_transfo}
In this section, we present the main ablation studies lead to understand the influence of the main hyperparameters and augmentation strategy on the Barlow Twins performance. For this study, the baseline is defined with a batch size of 128, $\lambda = 0.0051$ and a projector dimension of 4096. The study is led on a subset of ROI patches of pkgh-410, made of 5,000 patches per class.
\subsection{Hyperparameters}
\begin{figure}[htp]  
\begin{center}
        \includegraphics[scale = 0.25]{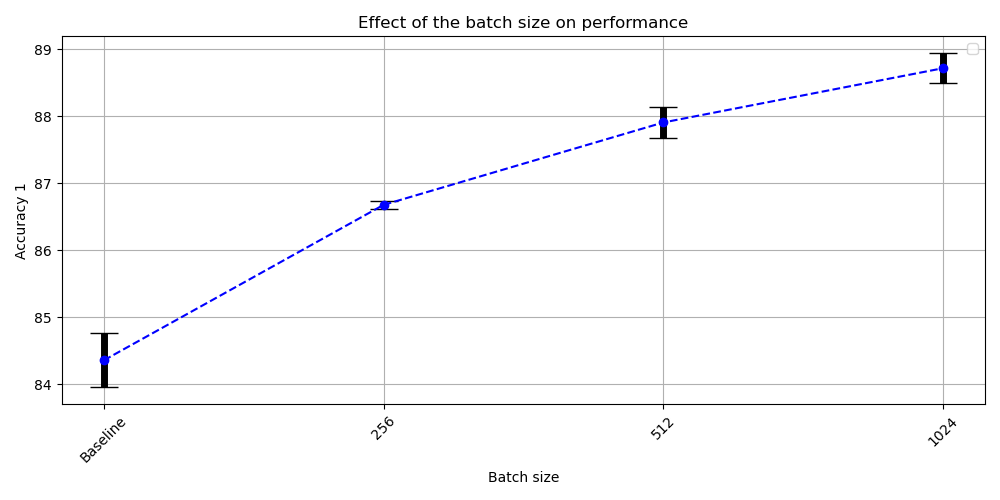}
\end{center}
    \caption{Influence of the batch size on the performance}
    \label{fig_BS}
\end{figure}

\begin{figure}[htp]  
\begin{center}
        \includegraphics[scale = 0.25]{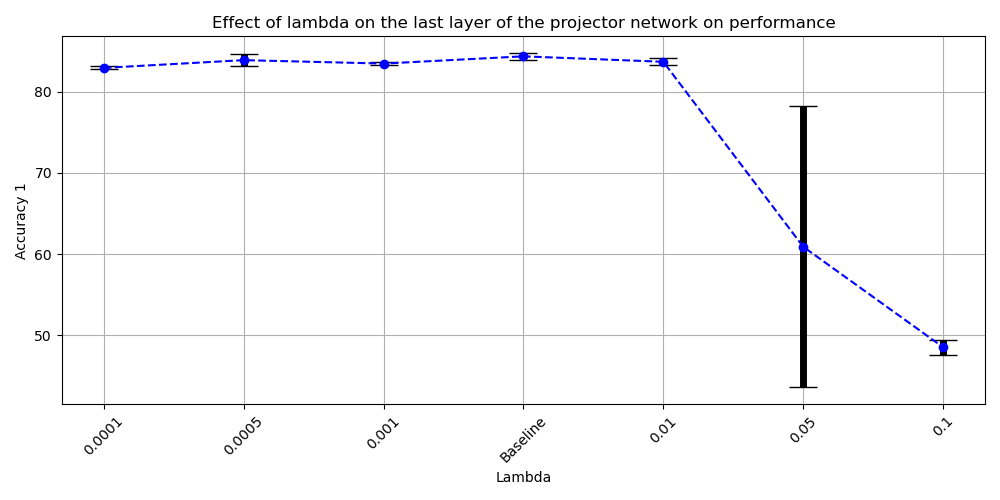}
\end{center}
    \caption{Influence of the $\lambda$ parameter on the model performance}
    \label{fig_lambda}
\end{figure}

\begin{figure}[htp]  
\begin{center}
        \includegraphics[scale = 0.25]{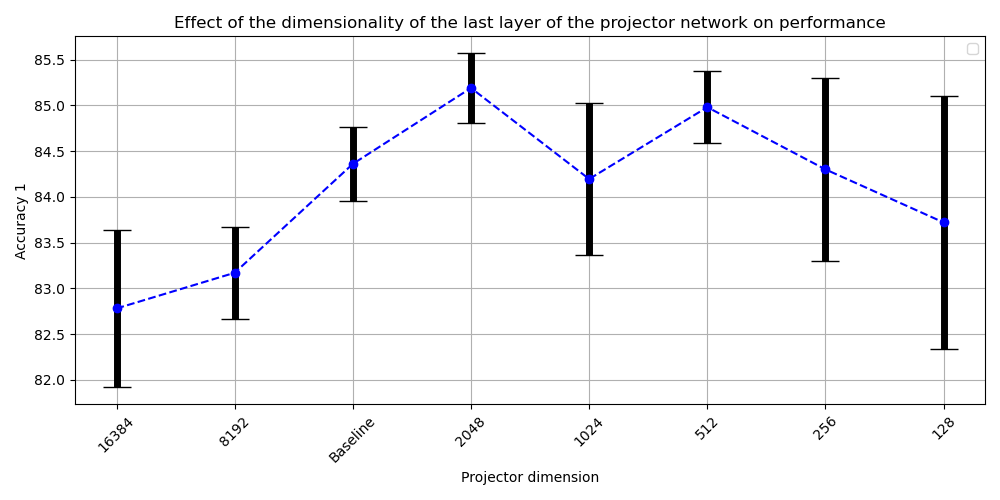}
\end{center}
    \caption{Influence of the projector dimension on the model performance}
    \label{fig_proj}
\end{figure}

The parameters studied here are the projector dimension, the batch size and the $\lambda$ (lambda) parameter in the loss function (Equation \ref{eq:l_bt}). Several key observations emerge from these experiments:
\newline \textbf{A larger batch size enhances performance:} in agreement with the original study lead by \cite{BT}, we conclude that a larger batch size, because it includes more batch statistics into the computation of the cross-correlation matrix, provides a more accurate loss and therefore leads to better representation. Figure \ref{fig_BS} shows the accuracy after linear evaluation of the model trained in Barlow Twins settings with different batch sizes.
\newline \textbf{The default $\lambda$ works well with pathology images:} as shown in Figure \ref{fig_lambda}, the value of the parameter does not influence much the training for $\lambda < 0.01$. However, a bigger $\lambda$ induces a high instability in the training. 
\newline \textbf{A smaller projector dimensions leads to better performance with the default augmentation strategy:} as shown in Figure \ref{fig_proj}, the better performance is obtained with a batch size of 2048 which comes in contradiction with the findings of the original paper \cite{BT}. This can be explained by the fact that this augmentation strategy causes the representations to overfit in a high-dimensional space. Moreover, a smaller projection head may result in underfitting, potentially eliminating meaningful data representations.
\newline \textbf{With the new augmentation strategy, a larger projection head is needed}: if we reproduce the ablation study on the projector dimension using the augmentation strategy proposed in this paper, we observe that the best results are obtained with a projector dimension of 8192. It means the representations learned with this new augmentation strategy require a higher dimensional space for proper representation.
\subsection{Augmentation strategy}
\begin{figure}[htp]  
\begin{center}
        \includegraphics[scale = 0.3]{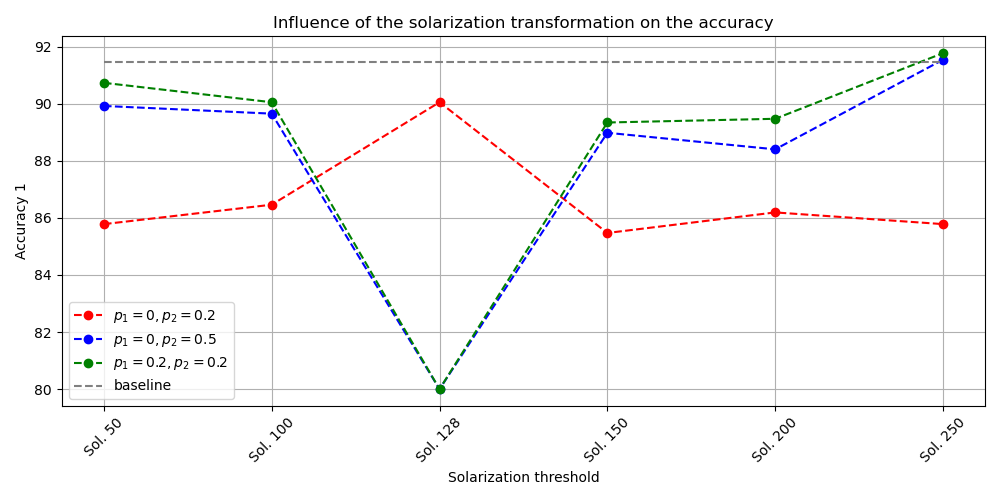}
\end{center}
    \caption{Influence of the solarization transformation on the model performance}
    \label{fig_sol}
\end{figure}

\begin{figure}[htp]  
\begin{center}
        \includegraphics[scale = 0.35]{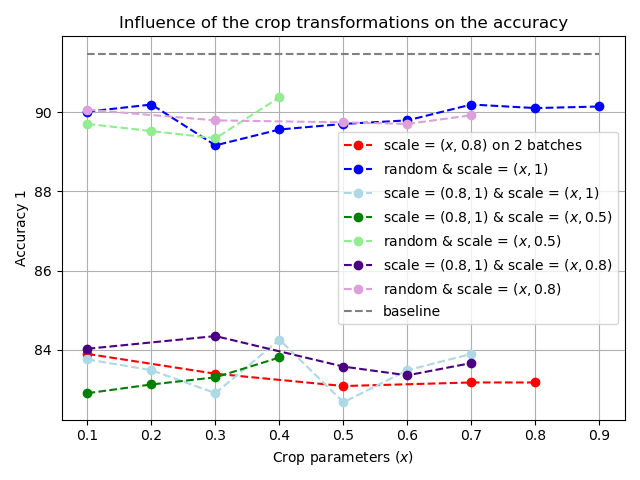}
\end{center}
    \caption{Influence of the sizes of the crops on the two batches on the model performance}
    \label{fig_crop}
\end{figure}

\begin{figure}[htp]  
\begin{center}
        \includegraphics[scale = 0.35]{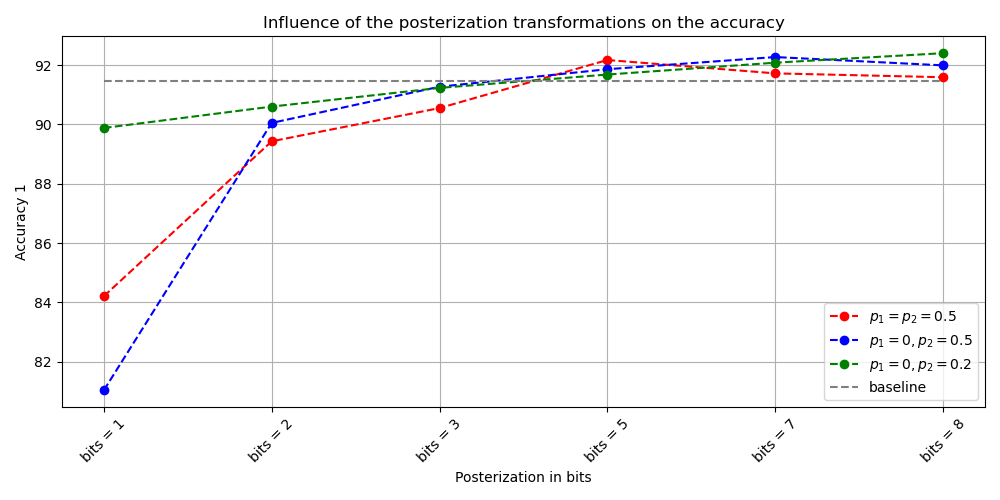}
\end{center}
    \caption{Influence of the posterization on the two batches on the model performance}
    \label{fig_post}
\end{figure}

\begin{figure}[htp]  
\begin{center}
        \includegraphics[scale = 0.35]{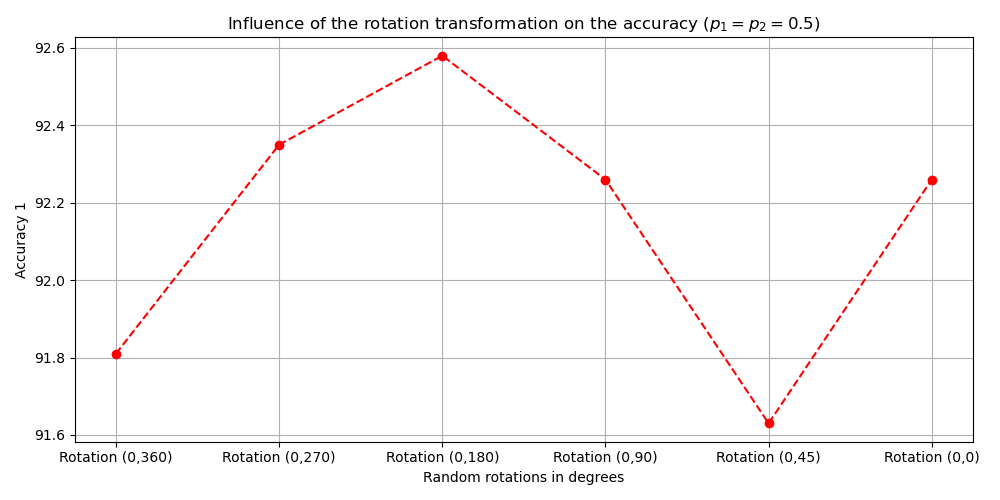}
\end{center}
    \caption{Influence of the rotation on the model performance}
    \label{fig_rot}
\end{figure}

\begin{figure}[htp]  
\begin{center}
        \includegraphics[scale = 0.35]{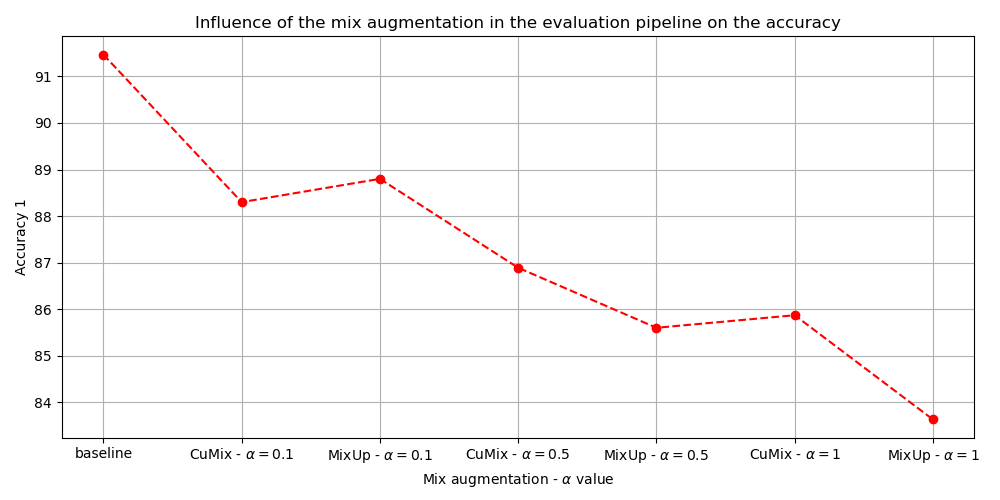}
\end{center}
    \caption{Influence of CutMix and MixUp in the evaluation phase on the test accuracy}
    \label{fig_mix_i}
\end{figure}

In this study, we first study the effects of removing specific data augmentations of the original augmentation strategies. Then, we study the thresholds used for Solarization, Colour Jitter, and Random Crops used in the original strategy before studying the potential of new transformations. Several key observations emerge from these experiments:
\newline \textbf{Grayscale and Gaussian blur harm the learning of good representations:} from Figure \ref{fig_transo_ab}, we understand that random grayscale and Gaussian blur, because they modify the meaningful colours and borders of the patches, destroy valuable information. Gaussian blur creates out-of-focus patches, which can harm representation learning in pathology.
\newline \textbf{A weak colour jitter benefits the training:} as supported by \cite{Benchmark}, a strong colour jitter results in unrealistic patches. One objective of the colour jitter is to expose the model to various colours and lighting conditions. It also prevents the model from overfitting to the specific colour characteristics of the training data and, therefore, is very useful when it comes to transfer learning, as the model will learn to generalize better and be able to adapt to diverse environments. In our case, the training dataset and the downstream dataset are the same, and therefore, the colour jittering is not very useful, even though it can provide more robustness to the training data. Applying a random weak colour jitter fosters the learning of meaningful representations by adding a weak and meaningful distortion.
\newline \textbf{A high solarization benefits the training}: removing solarization results in a modest boost in test accuracy. This observation aligns with previous research \cite{faryna2021tailoring} in which the authors highlight that solarization can harm pathology models because it produces unrealistic and unreliable images. However, as shown in Figure \ref{fig_sol}, a very high threshold of 250 colours the background or very thin tissue in blue and neutrophils, basophils, eosinophils and macrophages in dark blue, therefore emphasizing specific regions highly relevant to cancer diagnoses and reactive changes.
\newline \textbf{Barlow Twins does not benefit from different sizes of crops:} motivated by DINO settings \cite{DINO}, the possibility of using two different crop sizes for the input (local and global crops) is studied. Leveraging the symmetry of the Barlow Twins setup, the first batch is allocated to the global crop and the second batch to the local crop. We compare random crops using the final size or scale of the original patch. However, from Figure \ref{fig_crop}, we observe that having asymmetric crops does not benefit the training. Because different crops have different FoV, the model struggles to find useful data representations. Moreover, upsampling needs to be performed with a crop smaller than 224 × 224, and the data could become less realistic.
\newline \textbf{A high posterization can benefit the self-supervised training:} from Figure \ref{fig_post}, we observe that applying posterization with a limited number of bits does not benefit the training, whereas posterization on more than 5 bits outperforms the baseline (we note that the posterization on 8 bits stays within the range of the standard deviations of the baseline). The image on 7 bits is of slightly lower resolution than on 8 bits, which can contribute to catching high-frequency features, such as borders and contours.
\newline \textbf{Rotations leverage the non-canonical orientations of the pathology images:} we apply random rotations under different angles on the input patches. From Figure \ref{fig_rot}, we find that the best rotation angle is 180, which means the patches are rotated from any angle between 0 and 180. As flipping is already used, some rotations values could be redundant with the flips and lead back to the original orientations of the patch. This observation is supported by a recent paper introducing \texttt{HistoRotate} \cite{alfasly2023rotationagnostic}, a rotation method on the patch level to apply random rotation to one patch or to a crop of that patch and show great results.
\newline \textbf{Affine transformations enhance the self-supervised training}: using the same arguments as for the rotation, affine transformations leverage the non-canonical orientation of the patches and benefit the training. Moreover, affine transformation with
an angle between 0 and 45 and a translation factor of (0.5, 0.5) benefits the training and perform
better than a sole rotation between 0 and 180.
\newline \textbf{RandStainNA does not benefit the training:} RandStainNA, for Random Stain Normalization and Augmentation \citep{shen2022randstainna}, augments the images with more realistic stain styles and is a solution to bridge to stain augmentation and stain normalization. We observe that this augmentation does not benefit the training. However, our training and test data come from the same dataset and have the same stain. Therefore, this augmentation is not necessary in our case as we do not need to normalize the stain across multiple datasets and staining as they did in \cite{Benchmark}, for example. This augmentation could, however, be necessary if we want to transfer knowledge from the KGH dataset to another one (which could be H\&E stained, contrary to KGH, which is HPS stained). In this case, the training must be performed on the KGH dataset with RandStainNA augmentation. The results will be suboptimal but more transferrable.
\newline \textbf{MixUP \& CutMix are not appropriate for pathology downstream tasks:} Cutmix \citep{cutmix} is an augmentation technique particularly used for image classification tasks. It combines pairs of images and their labels to create new training examples. A crop from one patch is pasted onto another patch, and the labels are mixed according to the cut's proportions.  Similar to CutMix, MixUp operates by blending images, but it does so by taking a linear combination of pairs of images and their labels. Both methods operate on the images and the labels. If used during the pretraining, these methods perturb the learning of good representations. The highest test accuracy obtained was 85.96\% or 6 points below the baseline. One possible explanation is that the transformation alters the inherent structure within individual patches. However, in the context of pathology, the spatial arrangement of cells in the original image holds significance due to intercellular communication and potential interactions. Consequently, these augmentations introduce unrealistic data into the model. Despite concerns about realism, the demonstrated performance improvement for classification tasks suggests potential utility. Therefore, we propose exploring the application of these augmentations during the second phase, where patch classification is performed with the benefit of labelled data. Figure \ref{fig_mix_i} presents the test accuracy when these augmentations are applied to the training data during the evaluation phase. We observe that the baseline performs better. We conclude that as our patches can contain critical, subtle features that are key to diagnosis, applying MixUp or CutMix might obscure or distort these features, leading to misleading or uninterpretable images that do not resemble realistic pathological conditions.

\section{Supervised Training}
\label{sec_ab_sup}
To train the supervised ResNet-50 and Swin-Tiny, we performed an ablation study on the main hyperparameters: batch size (BS), learning rates, optimizer (opt), scheduler (sch) and weight decay (wd). The pretraining method (pr) was also evaluated. Table \ref{tab_sup_hp} presents the selected hyperparameters.
\begin{table}[h]
    \centering
    \footnotesize
    \begin{tabular}{c|c|c|c|c|c|c}
        \textbf{Encoder} & \textbf{BS} & \textbf{opt} & \textbf{LR} & \textbf{wd} & \textbf{sch} & \textbf{pr}\\\hline \hline
         ResNet-50 & 64 & Adam & 0.0001  & $1e^{-5}$ & none & ImageNet \\
         Swin-T & 64 & AdamW & 0.0001 & $1e^{-5}$ & none & ImageNet \\        
    \end{tabular}
    \caption{Main hyperparameters for the supervised training of ResNet-50 and Swin-T. LR for Learning Rate and BS for Batch Size}
    \label{tab_sup_hp}
\end{table}

Following this first study, we analyzed different data augmentation techniques and their impact on the training outcome. This study was made on a subset of pkgh-410 dataset and the ResNet-18 encoder for five epochs. In conclusion, and to reduce the overfitting of the models on the training data, the data augmentation techniques comprised:
\begin{itemize}
    \item random and vertical flips with a probability of 0.5;
    \item invertion with a probability of 0.5;
    \item rotation with a probability of 0.5 and \texttt{degrees=(0,90)}.
\end{itemize}

Then, the models were trained twice for 20 epochs and the model with the highest validation accuracy before overfitting was retained for further evaluation.
\section{Training Curves}
This section shows the training curves for Barlow Twins training, linear evaluation and CLAM training.
\label{sec_ab_curves}
\subsection{Barlow Twins}
\begin{figure*}[htp]  
\begin{center}
        \includegraphics[scale = 0.3]{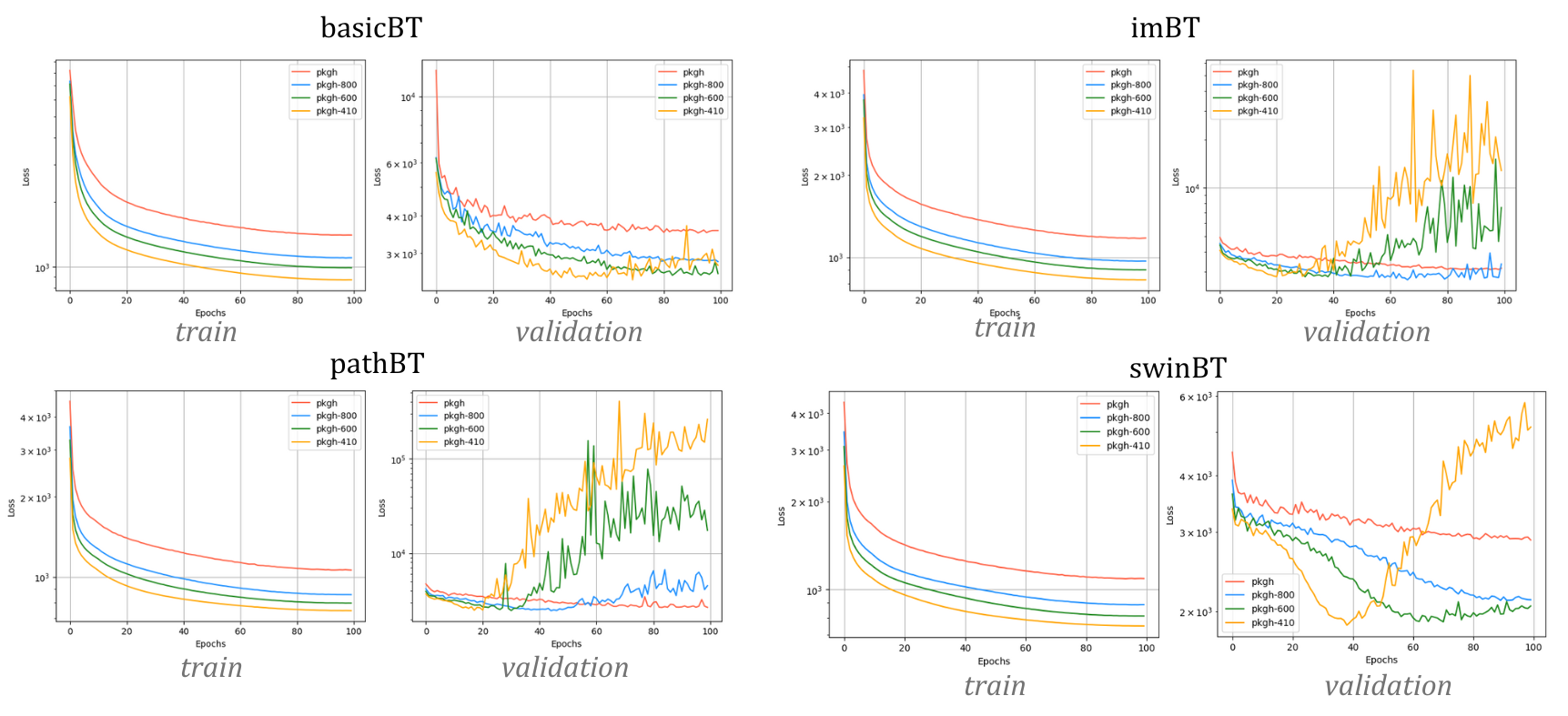}
\end{center}
    \caption{Training and validation losses for basicBT, imBT, pathBT and swinBT trained on pkgh, pkgh-800, pkgh-600 and pkgh-410. We can see that the training losses are decreasing over time however, the validation losses present with a different behaviour}
    \label{fig_BT_curves}
\end{figure*}

We follow the settings from Table \ref{tab_all_models} to train Barlow Twins. For the first phase, the encoders are trained on all patches (ROI, normal and non-ROI) for 100 epochs. Training and validation losses are reported in Figure \ref{fig_BT_curves}. From the training losses, we observe that the encoder is learning. Moreover, the validation losses in the case of Self-Supervised Learning reflect how well the model learns to understand the data structure through the pretext task and therefore, its measure of the quality of the learned representations is indirect. This is why it is more informative to evaluate the quality of the representations on a supervised downstream task: overfitting on the pretext task does not necessarily mean that the learnt representations are not of good quality.
\subsection{Linear Evaluation}
\begin{figure*}[htp]  
\begin{center}
        \includegraphics[scale = 0.45]{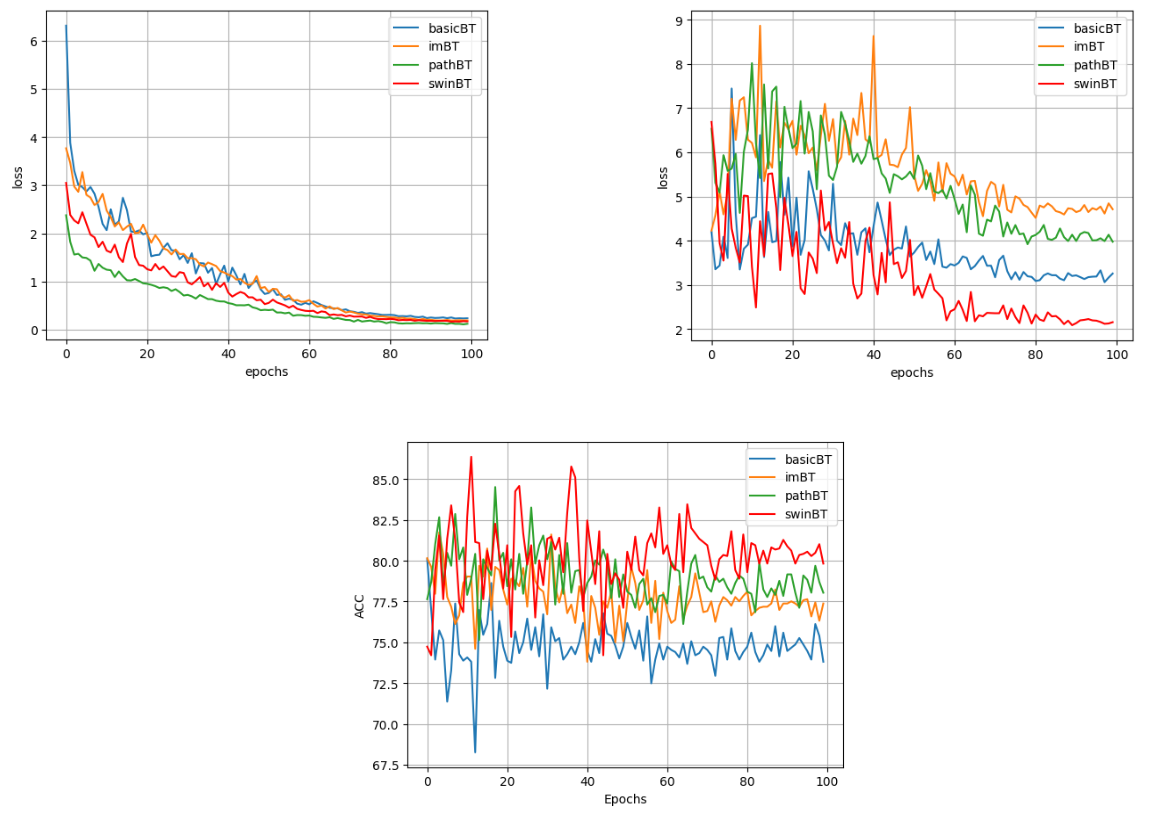}
\end{center}
    \caption{Training and validation losses (at the top from left to right) and validation accuracy (at the bottom) for basicBT, imBT, pathBT and swinBT trained on pkgh. }
    \label{fig_FC_curves}
\end{figure*}

We follow the original \href{https://github.com/facebookresearch/barlowtwins/tree/main}{code} to train the last linear layer. We observe that a lower initial learning rate than 0.3 provides highly fluctuating results and high overfitting. With an initial learning rate of 0.3, the loss fluctuates widely but decreases and the results are reproducible, showing that the model is learning. Figure \ref{fig_FC_curves} shows the training and validation losses (at the top, from left to right) and the validation accuracy (at the bottom) for the linear layers on top of basicBT, imBT, pathBT and swinBT models trained on pkgh.

\subsection{CLAM}
To train CLAM framework, we follow the original \href{https://github.com/mahmoodlab/CLAM}{code} which includes patch extraction, features extraction and MIL training. The original code was used for patch extraction, and therefore, artifact patches were extracted. However, the encoders were trained on the clean patches extracted following Section \ref{sec_ab_data}. Additionally, the CLAM patches extraction pipeline does not find any contour to process in some WSIs, where tissue samples are small and torn. This is a known issue from CLAM extraction pipeline. As a result, the number of slides used to train and test the models for the slide evaluation is modified. Table \ref{tab_clam_fail_count} presents the number of patches for the different classes and split.
\begin{table}[ht]
    \centering
    \small
    \begin{tabular}{|c|c|c|}
    \hline
        \textbf{class} & \textbf{train} & \textbf{test} \\\hline \hline
         \textbf{Normal} & 180 & 20  \\
         \textbf{HP} & 123 & 16 \\
         \textbf{SSLes} & 144 & 20 \\
         \textbf{TA} & 142 & 13\\
         \textbf{TVA} & 192 & 19 \\ \hline
         \textbf{TOTAL} & 781 & 88 \\\hline
    \end{tabular}
    \caption{Number of slides, extracted by CLAM, per class in the training and test sets}
    \label{tab_clam_fail_count}
\end{table}

\begin{figure*}[htp]
\begin{center}
        \includegraphics[scale = 0.3]{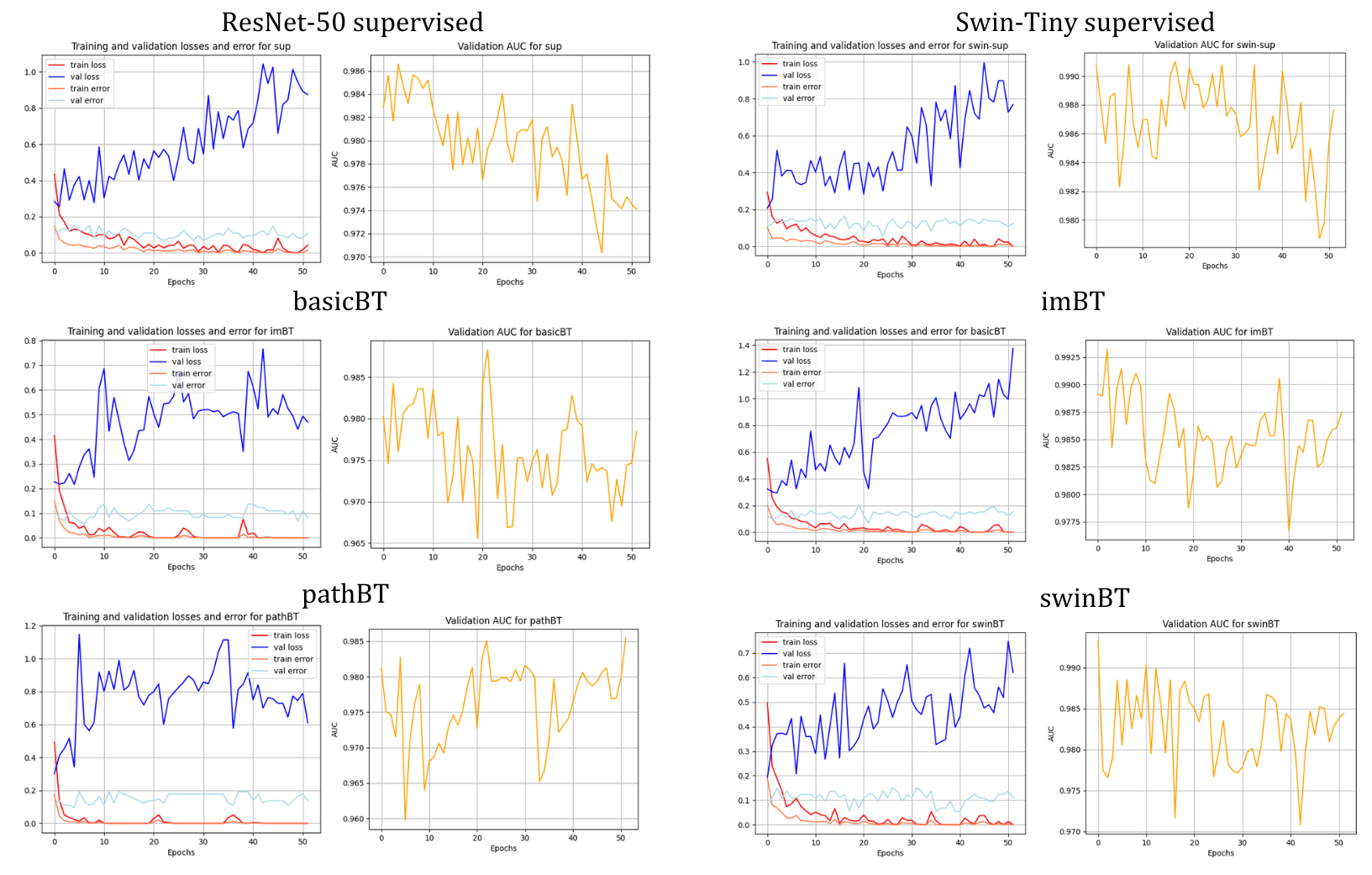}
\end{center}
    \caption{Training and validation losses and errors (in red and blue) and validation AUC (in yellow) for CLAM framework trained on top of basicBT, imBT, pathBT and swinBT encoders and trained on pkgh. The models all overfit but with pathBT for which we observe an overall increase of the validation AUC throughout the training}
    \label{fig_clam_curves}
\end{figure*}

Figure \ref{fig_clam_curves} shows the training and validation losses and errors as well as the validation AUC (yellow) of the training of the MIL model using the different pre-trained encoders. We can see that the models overfit, which is on par with current concerns regarding MIL method \cite{gadermayr2024multiple,yang2024mambamil}. However, we note that all models overfit regarding the validation AUC, but pathBT is the one for which the AUC increases throughout the training. This result highlights the pertinence of the novel augmentation for this model.
\section{Additional results and visualization}
\label{sec_ab_res}
\subsection{Detailed results for patch and slide evaluation}
Table \ref{tab_patch_acc} and Table \ref{tab_slide_c} present the accuracy, AUC and F1-score (if applicable) of the patch and slide classification taks. The polar graphs in Figures \ref{fig_patch_polar} and \ref{fig_slide_polar}.
\begin{table*}[htp]
    \tiny
    \centering
    \begin{tabular}{|c|c|c||c|c||c||c|c|}
        \hline
        \textit{Dataset} & \textbf{RN-50 sup.} & \textbf{Swin-T sup.} & \textbf{basicBT} & \textbf{imBT} & \textbf{benchBT} & \textbf{pathBT} & \textbf{swinBT} \\ \hline \hline
        \textbf{pkgh} & $81.85_{\pm0.05}$ & $82.67_{\pm0.75}$ & $78.77_{\pm1.87}$ & $81.32_{\pm0.6}$ & $75.86_{\pm0.09}$ & \textbf{$83.43_{\pm1.54}$} & \textbf{$86.15_{\pm0.33}$} \\
        &0.893 & 0.9119 & 0.9592 & 0.9618 & 0.8478 & \textbf{0.9652} & \textbf{0.9764} \\
        & \textit{} & & \textit{100} & \textit{100} & . & \textit{50} & \textit{100} \\ \hline 
        \textbf{pkgh-800} & $80.6_{\pm0.18}$ &$84.17_{\pm0.23}$ &  $81.27_{\pm0.57}$ & $81.3_{\pm0.14}$ & $75.9_{\pm1.94}$ & $83.84_{\pm0.05}$ & $86.47_{\pm2.4}$ \\
        & 0.909 & 0.9161 & 0.9668 & 0.9576 & 0.842 & \textbf{0.9676} & \textbf{\textcolor{red}{0.9822}} \\
        & & \textit{20} & \textit{100} & \textit{100} & . & \textit{100} & \textit{100} \\\hline
        \textbf{pkgh-600} & $81.43_{\pm0.14}$ & $83.17_{\pm0.9}$ &$80.54_{\pm0.09}$ & $82.26_{\pm1.03}$ & $74.9_{\pm0.33}$ & $81.84_{\pm0.49}$ & $84.17_{\pm1.6}$ \\
        & 0.91 & 0.8519 & 0.9613 & \textbf{0.9613} & 0.8424 & 0.9599 & \textbf{0.9728} \\
        & \textit{} & &\textit{100} & \textit{50} & . & \textit{100} & \textit{100}  \\ \hline 
        \textbf{pkgh-410} & $80.52_{\pm0.64}$& $80.86_{\pm}$& $80.06_{\pm}$ & $82.57_{\pm1.27}$& $74.03_{\pm0.37}$ & $80.4_{\pm0.8}$ & $82.37_{\pm0.65}$ \\
         & 0.8774 & 0.8965 & \textbf{0.9720} & 0.9649 & 0.8427 & 0.9515 & \textbf{0.9657}  \\
        & \textit{} & & \textit{100} & \textit{50} & . & \textit{100} & \textit{100} \\\hline 
    \end{tabular}
    \caption{Patch top-1 accuracy and AUC of the test set for all models and for all datasets. benchBT was trained for 200 ImageNet epochs \citep{Benchmark}. The number of epochs of Barlow Twins training after which the linear evaluation with the best performance was performed is in \textit{italic}. The \textbf{two best results} for each dataset are in bold. \textbf{\textcolor{red}{Best result overall}} is in \textbf{\textcolor{red}{red}}.}
    \label{tab_patch_acc}
\end{table*}

\begin{table*}[htp]
    \tiny
    \centering
    \begin{tabular}{|c|c|c||c|c||c||c|c|} 
    \hline
    \textit{Dataset} & \textbf{RN-50 sup.} & \textbf{Swin-T sup.} & \textbf{basicBT} & \textbf{imBT} & \textbf{benchBT} & \textbf{PathBT} & \textbf{SwinBT} \\ \hline \hline

    \textbf{pkgh} & 78.08 & 87.67 & 86.3 & 90.41 & 80.82 & \textbf{91.8} & \textbf{91.78}  \\
    & 0.9647 & 0.9807 & 0.9819 & 0.9854 & 0.9552 & \textbf{0.9862} & \textbf{0.9908}  \\
    & 76.89 & 87.86 & 86.29 & 90.29 & 78.86 & \textbf{91.8} & \textbf{91.79}  \\ \hline

    \textbf{pkgh-800} & 91.78 & \textbf{91.78} & \textbf{\textcolor{red}{94.52}} & 91.7 & 89.04 & 89.04 & 87.67 \\
    & 0.9935 & \textbf{\textcolor{red}{0.9951}} & {0.9894} & \textbf{0.9899} & 0.976 & 0.9812 & 0.9891 \\
    & 91.89 & \textbf{92.05} & \textbf{\textcolor{red}{94.51}} & 91.6 & 89.01 & 89 & 88.07 \\ \hline

    \textbf{pkgh-600} & 87.67 & 89.04 & 89.04 & \textbf{91.8} & & 89.04 & \textbf{{93.15}}  \\ 
    & 0.9856 & \textbf{0.9908} & 0.9842 & 0.9896 & . & 0.9812 & \textbf{{0.9932}} \\
    & 87.26 & 88.91 & 88.65 & \textbf{91.73} & & 89 & \textbf{{93.16}} \\ \hline 

    \textbf{pkgh-410} & \textbf{91.78} & 87.67 & 90.41 & \textbf{\textcolor{red}{95.89}} & & 87.67 & 89.04  \\
    & 0.9905 & 0.984 & \textbf{0.9941} & \textbf{\textcolor{red}{0.9966}} & . & 0.9819 & 0.992 \\
    & \textbf{91.73} & 87.74 & 89.62 & \textbf{\textcolor{red}{95.89}}  & & 87.24 & 89.11  \\ \hline 
    
    \end{tabular}
    \caption{Slide level accuracy, AUC and F1 score (from top to bottom) of the test slides for all models and all datasets. The two best results for each dataset are in bold.  \textcolor{red}{The two best results overall} are in \textcolor{red}{red}.}
    \label{tab_slide_c}
\end{table*}
\subsection{Confusion matrices}
\begin{figure*}[htp]  
\begin{center}
        \includegraphics[scale = 0.3]{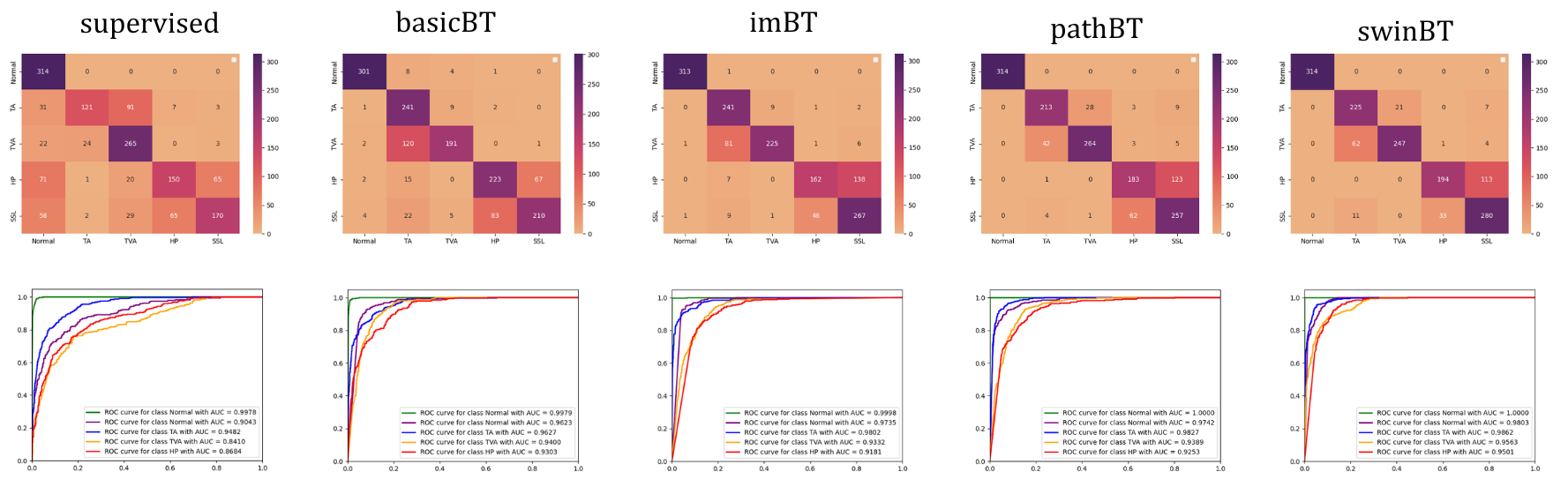}
\end{center}
    \caption{Confusion matrices and ROC curves for basicBT, imBT, pathBT and swinBT trained on pkgh. Here Sessile Serrated Lesions are abbreviated as SSL. We observe that the models have a very accurate ROC curve when it comes to Normal classification however, they struggle to differentiate HP and SSL, and TA and TVA, as the confusion matrices can emphasize.}
    \label{fig_CF_ROC}
\end{figure*}

Figure \ref{fig_CF_ROC} presents the confusion matrices and ROC curves for the supervised ResNet-50, basicBT, imBT, pathBT and swinBT. The first observation is that, even though the supervised baseline can separate the histology from the pathology patches (high ROC curve for Normal), it does not perform well to dissociate between the pathologies. The benchmark model can recognize HP and SSLe but struggles to recognize TA and TVA and does not separate histology well from pathology. The proposed models perform very well in recognizing histology from pathology. The swinBT obtains the best ROC curves. We remark that they struggle to separate HP from SSLe and TA from TVA even though imBT seems stronger in recognizing TVA from TA (almost no TVA are predicted as TA, but TA can be predicted as TVA). However, all models obtain better ROC curves for the classes TA/TVA than HP/SSLe. This can be explained by the fact that TVA is TA with more than 20\% of villous architecture. Therefore, TA structures can still have a bit of villous architecture. Additionally, SSLe is more common in the right colon, and HP in the left \citep{pickhardt2018natural}. By default, a growth in the right colon will be considered SSLe until proven otherwise. 

According to our expert pathologist, these results show that the models are sharp enough to detect this ambiguity. These results are highly valuable as the models catch these villous architectures that pathologists should consider. 
\subsection{CLAM heatmaps}
\begin{figure*}[htp]  
\begin{center}
        \includegraphics[scale = 0.7]{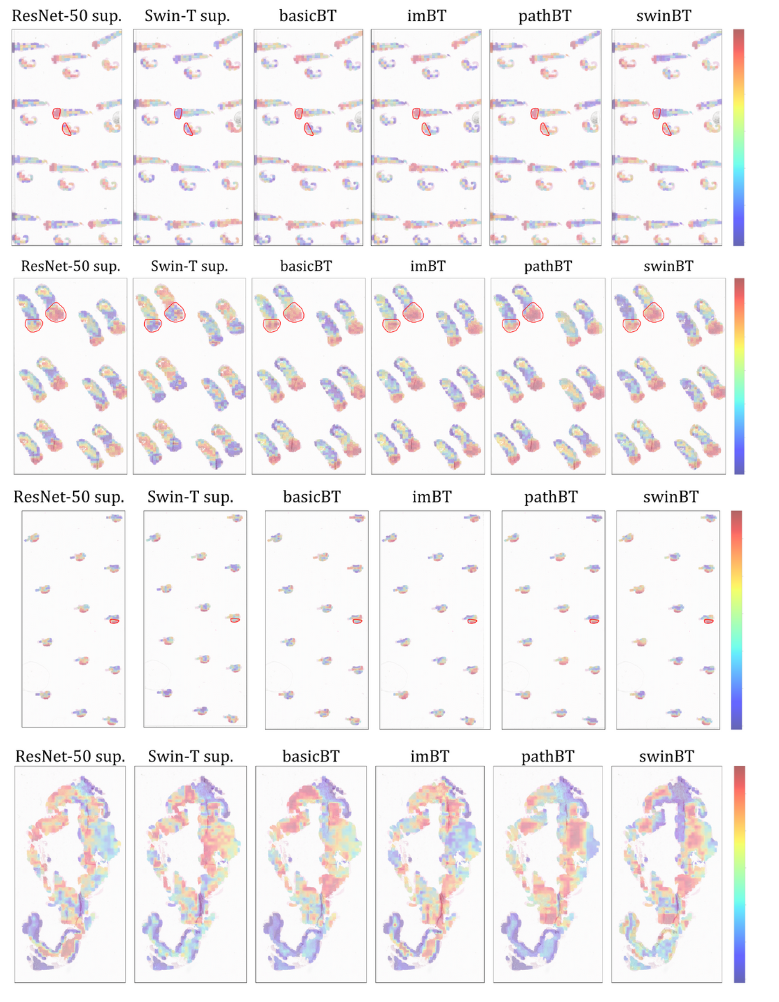}
\end{center}
    \caption{CLAM heatmaps for (from top to bottom) Hyperplastic Polyps, Sessile Serrated Lesions, Tubular Adenoma and Normal WSIs}
    \label{fig_heatmaps}
\end{figure*}

To complete the analysis of Section \ref{sec_clam_results}, sile additional heatmaps are displayed in Figure \ref{fig_heatmaps}. Overall, all the highest attention scores for the SSL methods map to the ROI from our expert pathologist. However, the patches with the highest attention scores are not always consistent for the supervised methods. Additionally, pathBT and swinBT provide more accurate heatmaps for the Hyperplastic Polyps (top) and Normal WSIs, we observe that basicBT and imBT are getting distracted by the tissue overlap regions, whereas pathBT and swinBT label them as non-relevant. pathBT and swinBT have similar relevant regions, and basicBT and imBT share similar diagnostically relevant regions as well. 

Example citation, See \citet{lamport94}.

\bibliographystyle{elsarticle-num} 
\bibliography{refs}

\end{document}